\documentclass[aps,prl,amsmath,twocolumn,groupedaddress,letterpaper,floatfix]{revtex4-1}
 \usepackage{graphicx,color}
 \usepackage{verbatim}
 \usepackage{amssymb}   % for math
 \usepackage{amsmath}
 \usepackage{amsfonts}
 \usepackage{mathdots}
 \usepackage{hyperref}
 \usepackage{epsfig}
 \usepackage{braket}
 \usepackage{bm}
 \DeclareMathOperator{\Tr}{Tr}
 \usepackage{soul} 
\usepackage{tikz}
\newcommand*\circled[1]{\tikz[baseline=(char.base)]{
		\node[shape=circle,draw,inner sep=0.3pt] (char) {#1};}} 
\usepackage{extarrows}
\usepackage{bm}
\usepackage{esint}
\usepackage{footnote}

\begin{document}
	\title{Majorana Zero Modes in the Lieb-Kitaev Model with Tunable Quantum Metric}

	\author{Xingyao Guo}\thanks{These authors contributed equally to this work.}
	\author{Xinglei Ma}\thanks{These authors contributed equally to this work.}
	\author{Xuzhe Ying}
	\author{K. T. Law}\email{phlaw@ust.hk}
	\affiliation{Department of Physics, Hong Kong University of Science and Technology, Clear Water Bay, Hong Kong, China} 	
	
	\date{\today}
	%%%%%%%%%%%%%%%%%%%%%%%%%%%%%%%%%%%%%%%%%%%%%%%%%%%%%%%%%%%%%%%%%%%%%%%%%%%%%%%%%
	\begin{abstract}  The relation between band topology and Majorana zero energy modes (MZMs) in topological superconductors had been well studied in the past decades. However, the relation between the quantum metric and MZMs has yet to be understood. In this work, we first construct a three band Lieb-like lattice model with an isolated flat band and tunable quantum metric. By introducing nearest neighbor equal spin pairing, we obtain the Lieb-Kitaev model which supports MZMs. When the Fermi energy is set within the flat band energy, the MZMs appear which are supposed to be well-localized at the ends of the 1D superconductor due to the flatness of the band. On the contrary, we show both numerically and analytically that the localization length of the MZMs is controlled by a length scale defined by the quantum metric of the flat band, which we call the quantum metric length (QML). The QML can be several orders of magnitude longer than the conventional BCS superconducting coherence length. When the QML is comparable to the length of the superconductor, the two MZMs from the two ends of the superconductor can hybridize and induce ultra long-range crossed Andreev reflections. This work unveils how the quantum metric can greatly influence the properties of MZMs through the QML and the results can be generalized to other topological bound states.
	\end{abstract}
	\maketitle
	
\begin{figure}[th]
		\centering 
		\includegraphics[width=1\linewidth]{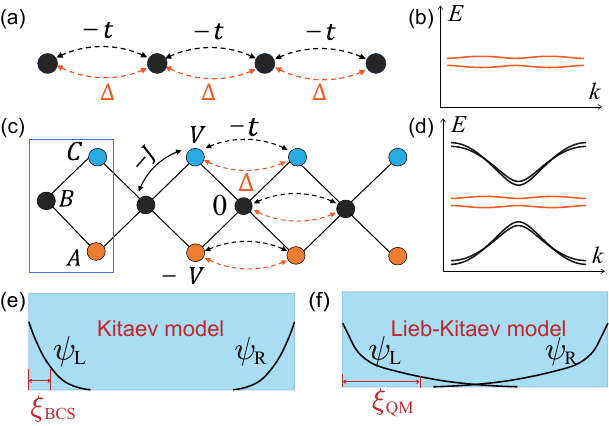}
		\caption{(a) and (c): Schematic illustrations of the real space single band Kitaev model  (a) and the three band Lieb-Kitaev model (c). In (a), $t$ denotes the nearest neighbor hopping and $\Delta$ is the nearest neighbor pairing. In (c): Each unit cell contains three lattice sites $A$,$B$ and $C$, respectively. $J$, $t$ and $\Delta$ are the nearest neighbor hopping, the next nearest neighbor hopping, and the nearest neighbor pairing, respectively. (b) and (d): Illustrations of the band structure of the Kitaev model (b) and the Lieb-Kitaev model (d). (e) and (f): Illustrations of the Majorana wavefunctions $\psi_{\text{L}}$ and $\psi_{\text{R}}$ of the Kitaev model (e) and the Lieb-Kitaev model (f). The spread of $\psi_{\text{L/R}}$ is controlled by $\xi_{\text{BCS}} \approx at/\Delta$ in the Kitaev model, and by the QML $\xi_{\text{QM}}$ in the Lieb-Kitaev model, respectively. }
		\label{figillustration}
	\end{figure}

	\emph{Introduction.}--- Majorana zero energy modes (MZMs) are non-Abelian excitations in topological superconductors \cite{PhysRevB.61.10267,AYuKitaev_2001, PhysRevLett.86.268,PhysRevB.73.014505,PhysRevLett.100.096407,PhysRevB.77.220501,PhysRevLett.101.120403,PhysRevB.79.161408,PhysRevLett.102.216403,PhysRevLett.102.216404,PhysRevLett.103.020401,PhysRevLett.103.237001,PhysRevLett.104.040502,PhysRevB.81.125318,PhysRevLett.105.077001,PhysRevLett.105.177002,PhysRevB.82.180516,PhysRevLett.105.227003,Alicea2010NonAbelianSA,PhysRevB.84.201105,PhysRevLett.112.037001}. Due to the non-Abelian nature of MZMs and their ability to store quantum information which are immune to local perturbations \cite{KITAEV20032,RevModPhys.80.1083,PhysRevX.6.031016}, the study of MZMs has been one of the most important topics in condensed matter physics in the past few decades \cite{RevModPhys.82.3045, RevModPhys.83.1057, Alicea_2012, Beenakkerreview}. As first pointed out by Read and Green \cite{PhysRevB.61.10267}, two-dimensional $p+ip$ superconductors which are characterized by nontrivial Chern numbers support chiral Majorana edge modes. The Chern number is defined as the sum of the Berry curvature of occupied quasiparticle states of the Bogoliubov-de Gennes (BdG) Hamiltonian.  Using a single band model with spinless $p$-wave pairing (Fig.~\ref{figillustration}(a)), Kitaev pointed out that one-dimensional topological superconductors support MZMs and the MZMs are localized at the two ends of the superconducting wires \cite{AYuKitaev_2001}. 
	
	After the above seminal works \cite{PhysRevB.61.10267, AYuKitaev_2001}, a large number of studies had contributed to the experimental realization and detection of MZMs \cite{doi:10.1126/science.1222360, rokhinson2012fractional, das2012zero, deng2012anomalous, nadj2014observation, albrecht2016exponential, jack2019observation, fornieri2019evidence, ren2019topological, wang2018evidence}. However, previous works were mostly focused on the topological aspects of MZMs, which are essentially connected to the Berry curvatures of the quasiparticle states. Interestingly, the Berry curvature is only one of the two aspects of the so-called quantum geometry of the quantum states \cite{cmp/1103908308,1984RSPSA.392...45B,Resta2010TheIS}. Given a Bloch state labeled by crystal momentum $\bm{k}$, we can construct the  quantum geometry tensor $\mathfrak{G}(\bm{k}) = \mathcal{G}(\bm{k}) - i\mathcal{F}(\bm{k})/2$ \cite{Resta2010TheIS}.  Here, the real part $\mathcal{G}$ is the quantum metric tensor of the Bloch states and the imaginary part $\mathcal{F}$ is the Berry curvature.  While the study of quantum metric effects has attracted much attention in recent years \cite{PhysRevLett.107.116801, CRPHYS_2013__14_9-10_816_0, PhysRevB.87.245103, PhysRevLett.112.166601, PhysRevB.90.165139, torma2015NC, PhysRevLett.117.045303, PhysRevB.95.024515, PhysRevA.97.033625, PhysRevB.102.201112, doi:10.1073/pnas.2106744118,PhysRevLett.126.156602,PhysRevB.104.L100501,PhysRevLett.127.170404,PhysRevResearch.3.L042018,PhysRevLett.128.087002,naturephysicsAhn,PhysRevLett.130.226001,PhysRevLett.132.026002,PhysRevLett.132.036001}, the relation between the quantum metric and the properties of the topological bound states has yet to be understood. This work is devoted to understand the connection between the quantum metric and the properties of the MZMs. 
	
\begin{figure*}[ht]
		\centering 
		\includegraphics[width=1\linewidth]{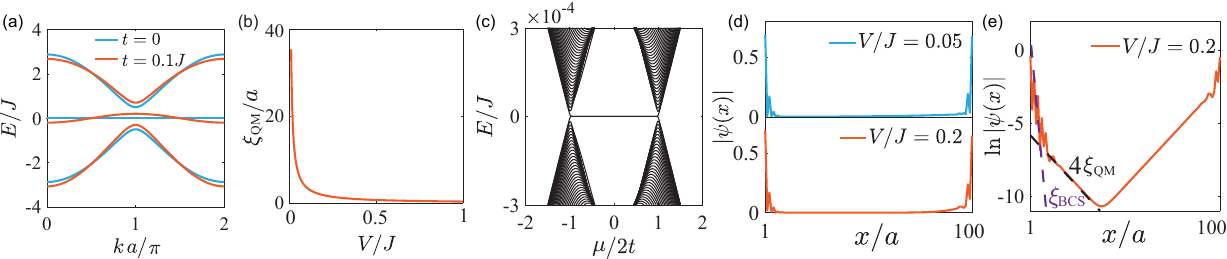}
		\caption{(a) The normal state energy spectrum with $t=0$ (blue line) and $t=0.1J$ (orange line), respectively.  $J=1$ and $V =0.2J$ are assumed in Eq.~(\ref{Eq_hnormal}). (b) QML as a function of $V/J$. (c) Energy levels of a lattice model of $H_{\text{BDG}}$ as a function of $\mu/2t$ with open boundary conditions. The parameters are $\mu=0$, $t = 3\times 10^{-4}J$, $\Delta=0.6t$ and $V=0.2J$. The zero energy modes appear in the topological regime $|\mu|< 2t$. (d) The wavefunction amplitudes of MZMs with different $V/J$. The length of the superconductor $L=100a$. The parameters are the same as in (c) except $V/J$. (e) $\ln{|\psi(x)|}$ is plotted for $|\psi(x)|$ in (d) with $V/J=0.2$. The short distance localization length is controlled by $\xi_{\text{BCS}}$ (dotted purple line) and the long distance localization length (dotted black line) is controlled by $4\xi_{\text{QM}}$. The numerical results are in perfect agreement with Eq.~(\ref{PsiL}). }
		\label{figmajorana1}
	\end{figure*}

	To study the quantum metric effects on MZMs, we introduce the Lieb-Kitaev model which supports MZMs and with tunable quantum metric. In the normal state, the model is a spinless Lieb-like lattice with three orbitals per unit cell (Fig.~\ref{figillustration}(c)). This lattice structure results in a (nearly) flat band between the two dispersive bands in the energy spectrum (Fig.~\ref{figmajorana1}(a)). Subsequently, we add the nearest neighbor intra-orbital pairings to the Lieb-like lattice to create a $p$-wave superconductor with MZMs, when the chemical potential is within the flat band energy. The resulting Bogoliubov quasiparticle bands are schematically illustrated in Fig.~\ref{figillustration}(d). The Majorana wavefunctions are illustrated in Fig.~\ref{figillustration}(f).
	
	There are three important results in this work. First, the quantum metric, which measures the quantum distance between two Bloch states \cite{cmp/1103908308,Resta2010TheIS}, indeed sets a length scale in real space which we call the quantum metric length (QML) $\xi_{\text{QM}}$ as defined in Eq.~(\ref{qmlength}) \cite{PhysRevLett.132.026002, hu2024anomalous}. The QML, defined as the average of the quantum metric over the Brillouin zone, governs the localization length as well as the quadratic spread of the Majorana wavefunctions for superconductors with (nearly) flat bands. Importantly, the QML is tunable and it can be orders of magnitude longer than the lattice length scale as illustrated in Fig.~\ref{figmajorana1}(b). Second, in flat band topological superconductors with long QML, the two MZMs from the two ends of the topological superconductor can hybridize with each other over a long distance even though the conventional BCS superconducting coherence length $\xi_{\text{BCS}}$ of the flat band superconductor is short. Here, $\xi_{\text{BCS}} \approx at/\Delta$ where $2t$ is the bandwidth, $\Delta$ is the pairing amplitude of the flat band and $a$ is the lattice constant. Third, the hybridization of MZMs can result in long range nonlocal transport processes such as crossed Andreev reflections (CARs) when two metallic leads are connected to the two MZMs separately \cite{PhysRevLett.103.237001, PhysRevLett.101.120403, PhysRevB.88.064509}. Remarkably, the CAR amplitudes can be comparable to the maximal theoretical value even when the separation of leads is several orders of magnitude longer than the $\xi_{\text{BCS}}$ of the flat band.
	
	\emph{Lieb-Kitaev model}.--- In this section, we introduce the Lieb-Kitaev model for the realization of  topological superconductors with tunable quantum metric. In the normal state of the Lieb-like lattice, the on-site energies of the $(A,B,C)$ orbitals are $(-V,0,V)$ respectively, as illustrated in Fig.~\ref{figillustration}(c). The nearest neighbor hopping amplitude is $J$. Additionally, a much smaller intra-orbital hopping $t$ is introduced (black dashed lines in Fig.~\ref{figillustration}(c)). Accordingly, the Hamiltonian in the Bloch basis $\hat{\bm{c}}(k) = \left(\hat{c}_A(k),\hat{c}_B(k),\hat{c}_C(k)\right)^{\text{T}}$ is written as
	\begin{equation}\label{Eq_hnormal}
		h(k) = (-2t\cos{(ka)}-\mu)\mathbb{I}_3 +\begin{pmatrix}
			-V& a_k & 0 \\
			a_k^*& 0& a_k^* \\
			0& a_k &  V \end{pmatrix}.
	\end{equation} 
	
	Here, $\mu$ is the chemical potential, $a_k = -J(1+ e^{ika})$ where $a$ is the lattice constant, and $\mathbb{I}_3$ is the identity matrix. Fig.~\ref{figmajorana1}(a) depicts the band structure of the model as defined in Eq.~(\ref{Eq_hnormal}). We focus on the (nearly) flat band with dispersion  $\epsilon_0=-2t\cos{(ka)} -\mu$, where $2t$ is the bandwidth of the flat band. When $t=0$, the band is exactly flat (blue line in Fig \ref{figmajorana1}(a)). The flat band is separated from two dispersive bands by an energy gap $E_{\text{g}}=|V|$.  The eigenstates of the flat band are
	\begin{equation}
		u_0(k)=(a_k, V, -a_k)^T/\sqrt{4J^2(1+\cos{(ka)})+V^2},
	\end{equation}	
which is essential for computing the quantum metric as well as constructing the Majorana wavefunctions as shown below.
	
	The quantum metric of a state with momentum $k$ of the flat band is defined as $\mathcal{G}(k) =  \bra{\partial_ku_0(k)}\left[\mathbb{I} -\ket{u_0(k)}\bra{u_0(k)} \right]\ket{\partial_ku_0(k)}$ which has the dimension of length-squared. The QML $\xi_{\text{QM}}$ is defined as the Brillouin zone averaged quantum metric:
	\begin{equation} \label{qmlength}
		\xi_{\text{QM}} \equiv \int_0^{2\pi/a}\mathcal{G}(k) \frac{dk}{2\pi}\overset{\frac{V}{J}\rightarrow0}{\longrightarrow} \frac{\sqrt{2}}{4}\frac{J}{V}a. 
	\end{equation}
	The length scale $\xi_{\text{QM}}$ is particularly important for exactly flat bands with $t=0$ and vanishing Fermi velocity. In this case, the conventional length scales such as the Fermi wavelength is not well-defined and the BCS coherence length $\xi_{\text{BCS}} \approx a t/\Delta$ is zero for flat bands. As we show below, the QML $\xi_{\text{QM}}$ is still a dominant length scale which governs the spread of the Majorana wavefunctions in topological superconductors when $\xi_{\text{QM}}$ is longer than $\xi_{\text{BCS}}$. Moreover, for the Lieb-like lattice, the $\xi_{\text{QM}}$ is tunable by changing $V/J$. Fig.~\ref{figmajorana1}(b) shows that $\xi_{\text{QM}}$ is divergent when $V/J$ approaches $0$. With the tunable QML, the Lieb-like lattice is an ideal model for studying the interplay between the topology and quantum metric.
	
	To realize MZMs, we introduce intra-orbital pairing with amplitude $\Delta$ between sites from adjacent unit cells, indicated by the red dashed line in Fig.~\ref{figillustration}(c). The resulting BdG Hamiltonian of the Lieb-Kitaev model is
	\begin{equation} \label{Eq_BdG_majorana}
		H_{\text{BdG}} = \sum_{k} \hat{\Psi}^\dagger(k) \begin{pmatrix}h(k)& -i2\Delta \sin{(ka)}
			\mathbb{I}_3\\
			i2\Delta \sin{(ka)} \mathbb{I}_3 &  -h^*(-k)\end{pmatrix}
		\hat{\Psi}(k), 
	\end{equation}
    where $ \hat{\Psi}(k) =  (\hat{\bm{c}}(k),\hat{\bm{c}}^\dagger(-k) )^{\text{T}}$. Fig.~\ref{figmajorana1}(c) shows the energy levels of a finite size system with open boundary conditions within the energy window $E\in[-t,t]$.  We observe that zero energy modes exist when the chemical potential lies in the region $|\mu|<2t$. The topological phase is characterized by the Z2 number $ Q = \operatorname{sign}(\text{Pf}[i\tilde{H}(k=0)] \text{Pf}[i\tilde{H}(k=\frac{\pi}{a})]$ \cite{AYuKitaev_2001, PhysRevB.78.195125}.  As shown in the Supplemental Material \cite{NoteX}, in cases of $|\mu|<2t$, we have $\mathcal{Q}=-1$, which corresponds to the topologically nontrivial regime.  
	
	Fig.~\ref{figmajorana1}(d) depicts the Majorana wavefunctions of the models with two different values of $V/J$ and $\Delta=0.6t$ such that the band is extremely flat. With a larger quantum metric (smaller $V/J$), the Majorana wavefunctions can penetrate deeper into the bulk of the flat band superconductor. The localization length of the Majorana wavefunctions is indeed much longer than $t/\Delta$ which is different from the single band Kitaev model. The asymmetry of the Majorana wavefunctions from the two ends of the superconductor originates from the  inversion symmetry breaking of the underlying lattice. To quantify the spread of the Majorana wavefunctions, we plot $\ln{|\psi(x)|}$ versus position $x$ in Fig.~\ref{figmajorana1}(e). Here, $\psi(x)$ is the wavefunction of the fermionic mode which includes the Majorana wavefunctions from the left and the right boundaries.  For the Majorana modes of the left boundary, for example, there are two different decay modes in short and long distances as shown in Fig.~\ref{figmajorana1}(e). At a relative short distance away from the left boundary, the Majorana wavefunction decays as $ e^{-x/\xi_{\text{BCS}}}$ (purple dashed line in Fig.~\ref{figmajorana1}(e)), where $\xi_{\text{BCS}} = -2a/\ln{\left(\frac{t-\Delta}{t+\Delta}\right)}$ \cite{leumer2020exact}. For $t \gg \Delta$, we have $\xi_{\text{BCS}} \approx a t/\Delta$ and we call this length the BCS coherence length. However, at larger $x$, a different decay behavior takes over and the wavefunction decays as $e^{-x/4\xi_{\text{QM}}}$ (black dashed line in Fig.~\ref{figmajorana1}(e)), where $\xi_{\text{QM}}$ is the QML defined in Eq. (\ref{qmlength}). In the next section, we will show analytically how the QML emerges in the Majorana wavefunctions.
	
	%%%%%%%%%%%%%%%%%%%%%%%%%%%%%%%%%%%%%%%%%%%%%%%%%%%%%%%%%%%%%%%%%%%%%%%%%%%%%%%%%%%%%%%%%%%%%%%%%%%%%%%%%%%%%%%%%%%%%%%%%%%%%%%%%%%%%%%%%%%%%%%%%%%%%%%%%%%%%%%%%%
	\emph{Wavefunctions, localization length and quadratic spread of MZMs.}--- To begin with, we consider the multiband Hamiltonian $H = H_0+H_1$, where $H_0$ is the lattice representation of Eq.~(\ref{Eq_BdG_majorana}) with $N$ sites and a periodic boundary condition. The perturbation $H_1$ removes the hopping and pairing between the first site $1$ and the last site $N$ of $H_0$ and the addition of $H_1$ results in a Hamiltonian with an open boundary condition \cite{PhysRevB.83.125109}. For $V\gg t \simeq \Delta$ and $|\mu|<2t$, the two isolated quasi-particle bands labeled by $n=\pm$ respectively are close to the Fermi energy and far away from other bands. The eigenstates of the $n=\pm$ bands are denoted by $g_{n}(k)$ where $H_{\text{BdG}}(k)  g_{n}(k) = \varepsilon_{n}(k) g_{n}(k) $. The eigenstate $\psi(x)$ which contains both of the left Majorana mode $\psi_{\text{L}}(x)$ and the right Majorana mode $\psi_{\text{R}}(x)$ can be expressed in a self-consistent way as:
   \begin{equation}\label{Mwf}\begin{aligned}
   \psi(x) 
   &= \psi_{\text{L}}(x) + i\psi_{\text{R}}(x) \\
   &= -G^p(x,Na;E) U_{01}\psi(a) - G^p(x,a;E) U_{01}^\dagger \psi(Na),\end{aligned}
  \end{equation}
  where 
  \begin{equation} \label{projectedGF}
  G^p(x_j,x_{j^\prime};E) = \frac{1}{N}\sum_{n=\pm} \sum_{k} \frac{g_{n}(k) g^\dagger_{n}(k)}{E-\varepsilon_{n}(k)}e^{ik(x_j-x_{j^\prime})}  
  \end{equation}
  is the projected Green function where $j/j'$ is the site index and $E=0$ for the MZMs. The operator $U_{01}$ is the coupling matrix between adjacent unit cells. The details of the calculations for the Majorana wavefunctions are presented in the Supplemental Material \cite{NoteX}.  
  
  Away from the left boundary (the first site) and by setting $\mu=0$ for simplicity, the left Majorana wavefunction can be written as  
   \begin{equation}
  	\psi_{\text{L}}(x_j)= A_{\text{L}}^{\text{QM}} e^{-\frac{(j-1)a}{4\xi_{\text{QM}}}} +
  	A_{\text{L}}^{\text{BCS}} e^{-\frac{(j-1)a}{\xi_{\text{BCS}}}}. \label{PsiL}
  \end{equation}
  
\begin{figure}[h]
  	\centering 
  	\includegraphics[width=1\linewidth]{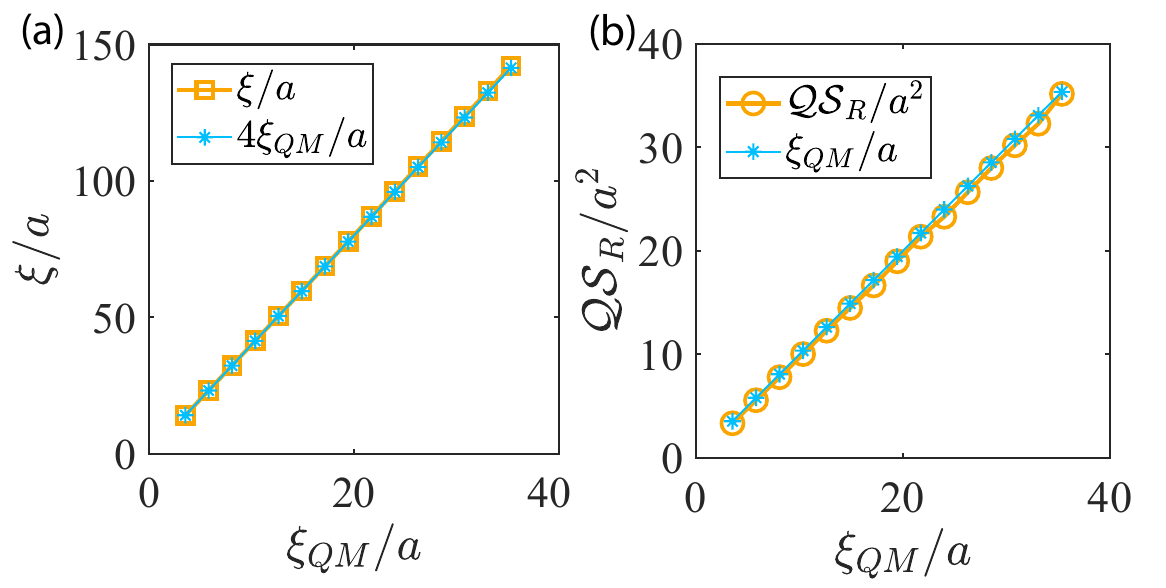}
  	\caption{(a) A comparison between the localization length of a MZM extracted numerically $\xi$, and $4\xi_{\text{QM}}$. (b) The quadratic spread $\mathcal{QS}$ of MZMs as a function of QML $\xi_{\text{QM}}$. The results are in agreement with Eq.~(\ref{QS}). The parameters of the lattice model for both figures are $\mu=0$, $t =5\times 10^{-4}J$, $\Delta = 4.95\times 10^{-4}J$ and $V/J \in [0.01, 0.1]$. }
  	\label{figmajorana2}
\end{figure}

  Here, $A_{\text{L}}^{\text{QM}}$ and  $A_{\text{L}}^{\text{BCS}} $ are the amplitudes of two parts of the wavefunction with different localization lengths $4\xi_{\text{QM}}$ and $\xi_{\text{BCS}}$ respectively. The localization lengths are determined by the poles of  $g_{n}(k) g^\dagger_{n}(k)/\varepsilon_{n}(k)$ in the complex plane. Physically, $\xi_{\text{BCS}} =  -2a/\ln{\left(\frac{t-\Delta}{t+\Delta}\right)}$ originates from the dispersion of quasi-particle bands $\varepsilon_\pm (k)$. This decay length is the same as the one in the single band Kitaev model with bandwidth $2t$ and pairing potential $\Delta$ \cite{leumer2020exact}. Importantly, an extra pole of the Bloch wavefunctions gives rise to a decay length of $4\xi_{\text{QM}}$ for the $A_{\text{L}}^{\text{QM}}$ component of the wavefunction. When $4\xi_{\text{QM}} \gg \xi_{\text{BCS}}$, the QML $\xi_{\text{QM}}$ dominates the long range behavior of the Majorana wavefunction.  The $A_{\text{L}}^{\text{BCS}}$ component has different amplitudes for the even or odd lattice sites, which explains the oscillation of MZMs' wavefunction as a function of lattice site. A similar expression for the wavefunction localized near the right boundary (the Nth-site) $\psi_{\text{R}}(x_j)$ is shown the Supplemental Material \cite{NoteX}. To compare the analytical results with the numerical results, the long distance localization length of MZMs $\xi$ is extracted numerically (orange squared line) and it matches the analytical values of $4\xi_{\text{QM}}$ (blue stared line) perfectly, as shown in Fig.~\ref{figmajorana2}(a). 
  
Besides the localization length, the spread of a wavefunction can also be characterized by the quadratic spread which was used to measure the size of Wannier states \cite{PhysRevB.56.12847}. The quadratic spread of the right Majorana can be evaluated as $\mathcal{QS}_{\text{R}} \equiv	\sum_{x=a}^{Na} (x-x_{\text{R}})^2 |\psi_{\text{R}}(x)|^2$, and $x_{\text{R}}$ is the position of the right boundary. In the limit of small $V/J$, vanishing $\xi_{BCS}$ with $t\approx \Delta$ and $\mu=0$, to order $\mathcal{O}(V/J)$, we have
\begin{equation}
		\mathcal{QS}_{\text{R}} = \frac{a}{2\pi} \int_0^{2\pi/a} \mathcal{G}(k)dk  = a\xi_{\text{QM}}.  \label{QS}
\end{equation}

%The quadratic spread of the Majorana wavefunctions can be evaluated as $\mathcal{QS} =\mathcal{QS}_{\text{L}}+\mathcal{QS}_{\text{R}} = \int (x-x_{\text{L}})^2 |\psi_{\text{L}}(x)|^2 dx 	+ \int (x-x_{\text{R}})^2 |\psi_{\text{R}}(x)|^2 dx $, and $x_{\text{L(R)}}$ is the position of the left (right) boundary. In the limit of small $V/J$, $t\approx \Delta$ and $\mu=0$, to order $\mathcal{O}(V/J)$, we have
%  \begin{equation}
%  	\mathcal{QS}= \frac{a}{2\pi} \int_0^{2\pi/a} \mathcal{G}(k)dk  = a\xi_{\text{QM}}.  \label{QS}
%  \end{equation}

  The analytical results are also in agreement with the numerical results as shown in Fig.~\ref{figmajorana2}(b). This is one of the key result of this work as it connects the quantum metric with the spread of the Majorana wavefunctions. The details of the derivation for Eq.~(\ref{QS}) are given in the Supplemental Material \cite{NoteX}.  
  
 %%%%%%%%%%%%%%%%%%%%%%%%%%%%%%%%%%%%%%%%%%%%%%%%%%%%%%%%%%%%%%%%%%%%%%%%%%%%%%%%%%%%%%%%%%%%%%%%%%%%%%%%%%%%%%%%%%%%%%%%%%%%%%%%%%%%%%%%%%%%%%%%%%%%%%%%%%%%%%%%%%
	\emph{Long range crossed Andreev reflection.}--- In this section, we show that a long quantum metric length can induce long range nonlocal transport when two leads are coupled to the two MZMs separately. In particular, the CAR probability can nearly reach the maximal theoretical value even though the separation of the two leads is several orders of magnitude longer than the conventional localization length of the Majorana modes $\xi_{\text{BCS}}$ in the one band Kitaev model.
		
\begin{figure}[h] 
		\centering 
		\includegraphics[width=1\linewidth]{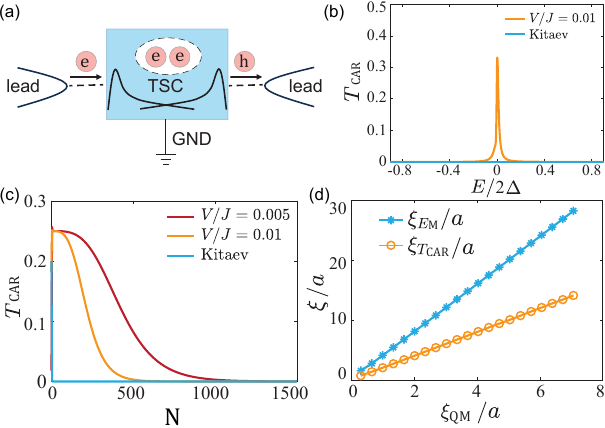}
		\caption{ (a) A schematic plot of a normal lead/topological superconductor (TSC)/normal lead device. Two leads couple to the two MZMs separately. In a CAR process, an electron from the left lead is reflected as a hole in the other lead. (b) The CAR probability versus bias voltage $E=eV$ for the single band Kitaev model (blue line) and the Lieb-Kitaev model (yellow line). Parameters $J=1$, $t = 1\times 10^{-4}J$, $\Delta =0.8t$, $\mu=0.2t$ and $L = 100a$. $V/J=0.01$ is set for Lieb-Kitaev model. The same $\mu$, $t$ and $\Delta$ are set for the single band Kitaev model. (c) The CAR probability at zero bias $E=0$ versus device length $L=Na$ for the Kitaev model (blue line) and the Lieb-Kitaev model with $V/J =0.005$ (red line) and $V/J =0.01$ (yellow line), respectively. (d) The long distance behavior of the hybridization energy $E_{\text{M}}$ of the MZMs and the CAR probability $T_{\text{CAR}} $ as a function of the length of the superconductor can be written as $E_{\text{M}}\propto e^{-Na/\xi_{E_{\text{M}}}}$ and  $T_{\text{CAR}} \propto e^{-Na/\xi_{T_{\text{CAR}}}}$. Here, we show that $\xi_{E_{\text{M}}} = 4\xi_{\text{QM}}$ and $\xi_{T_{\text{CAR}}} = 2\xi_{\text{QM}}$. The  parameters in (c) and (d) are the same as in (b) except $L$.} 
		\label{fig_CAR}
	\end{figure}

	Considering a device shown in Fig.~\ref{fig_CAR}(a), two normal metal leads are attached to two sides of the topological superconductor. A CAR process happens when an incoming electron from one lead is reflected as a hole in the other lead, leading to the formation of a Cooper pair in the grounded superconductor \cite{PhysRevLett.103.237001,PhysRevLett.101.120403,PhysRevB.88.064509}. Due to the quantum metric induced spread of the Majorana modes as discussed above, the coupling between Majorana modes can be significant in a long topological superconducting wire. We expect that the coupled Majorana modes can mediate long range CARs as shown below.
	
	To be more specific, we perform recursive Green function calculations \cite{datta1997electronic} to study the CAR probability $T_{\text{CAR}}$ for both the single band Kitaev model and the Lieb-Kitaev model. The blue line in Fig.~\ref{fig_CAR}(c) shows that the CAR signal only survives in a short wire with a few tens of lattice sites for the single band Kitaev model (Fig.~\ref{figillustration}(a)). The $T_{\text{CAR}}$ diminishes quickly once the length of the superconductor increases as the MZMs cannot couple to each other due to the short localization lengths of the MZMs in the single band Kitaev model $\xi_{\text{BCS}}$. 
	
	In sharp contrast, as indicated by the red and yellow lines in Fig.~\ref{fig_CAR}(b)-(c), the CAR probability is dramatically enhanced in the Lieb-Kitaev model with large quantum metric. The $T_{\text{CAR}}$ is most significant at low bias, as shown in Fig.~\ref{fig_CAR}(b). When the energy of the incoming electron is close to the energy of the fermionic mode formed by the hybrization of the MZMs, a large CAR probability, which is near the maximal theoretical value of 0.5, is possible \cite{PhysRevLett.101.120403} . Fig.~\ref{fig_CAR}(c) depicts $T_{\text{CAR}}$ versus the length of the superconductor at zero bias (red and yellow lines). We find that $T_{\text{CAR}}$ can be large when the separation of the leads is comparable to the QML even when the QML is several orders of magnitude longer than $at/\Delta$. It is striking that the CAR amplitude remains finite up to thousands of sites in cases of larger QML (red line in Fig.~\ref{fig_CAR}(c)) when $t/\Delta \approx 1$.
	
	A more careful analysis shows that at low voltage bias, the CAR probability is closely related to the coupling between the two MZMs. The strength of coupling between the MZMs is characterized by the hydridization energy $E_{\text{M}}$. As shown in Fig.~\ref{fig_CAR}(d), we found that the hydridization energy is proportional to the length of the topological superconductor such that $E_{\text{M}}\propto e^{-Na/4\xi_{\text{QM}}}$. Accordingly, the CAR probability can be expressed as $T_{\text{CAR}}\propto e^{-Na/2\xi_{\text{QM}}}$ (yellow line in Fig.~\ref{fig_CAR}(d)).

	%%%%%%%%%%%%%%%%%%%%%%%%%%%%%%%%%%%%%%%%%%%%%%%%%%%%%%%%%%%%%%%%%%%%%%%%%%%%%%%%%%%%%%%%%%%%%%%%%%%%%%%%%%%%%%%%%%%%%%%%%%%%%%%%%%%%%%%%%%%%%%%%%%%%%%%%%%%%%%%%%%
	\emph{Discussion.}--- In this work, we construct the Lieb-Kitaev model to study the effect of quantum metric on MZMs. It is shown that the localization length as well as the quadratic spread of the Majorana wavefunctions are controlled by the QML in a flat band topological superconductor. Importantly, the Majorana wavefunctions can spread far away from the boundaries when the QML is long. The two MZMs can couple to each other when the QML is comparable to the length of the flat band topological superconductor. The coupling of MZMs can induce long range CARs when two leads are coupled to the two ends of the topological superconductors. It is important to note that the Lieb-Kitaev model proposed only involves the nearest neighbor hopping and pairing. Therefore, quasi-one-dimensional moir\'e materials, which can be described by Lieb-like lattice \cite{PhysRevB.99.195455} in the normal state, can possibly be used to realize the Lieb-Kitaev model. Importantly, the QML is defined by the wavefunctions of the normal state. Therefore, the conclusions of this work can be generalized to describe topological bound states in other topological materials.
	 
	%%%%%%%%%%%%%%%%%%%%%%%%%%%%%%%%%%%%%%%%%%%%%%%%%%%%%%%%%%%%%%%%%%%%%%%%%%%%%%%%%%%%%%%%%%%%%%%%%%%%%%%%%%%%%%%%%%%%%%%%%%%%%%%%%%%%%%%%%%%%%%%%%%%%%%%%%%%%%%%%%%
	\emph{Acknowledgements.}--- We thank Shuai Chen, Adrian Po and Bohm-Jung Yang for inspiring discussions. K. T. L. acknowledges the support of the Ministry of Science and Technology, China, and Hong Kong Research Grant Council through Grants No. 2020YFA0309600, No. RFS2021-6S03, No. C6025-19G, No. AoE/P-701/20, No. 16310520, No. 16310219, No. 16307622, and No. 16309718.
	%%%%%%%%%%%%%%%%%%%%%%%%%%%%%%%%%%%%%%%%%%%%%%%%%%%%%%%%%%%%%%%%%%%%%%%%%%%%%%%%%%%%%%%%%%%%%%%%%%%%%%%%%%%%%%%%%%%%%%%%%%%%%%%%%%
	\bibliographystyle{apsrev4-1}
\let\oldaddcontentsline\addcontentsline% Store \addcontentsline
\renewcommand{\addcontentsline}[3]{}% Make \addcontentsline a no-op

\let\addcontentsline\oldaddcontentsline% Restore \addcontentsline

\clearpage
\onecolumngrid
\begin{center}
	\textbf{\large Supplemental Material for ``Majorana Zero Modes in the Lieb-Kitaev Model with Tunable Quantum Metric"}\\[.2cm]
	Xingyao Guo$^{*}$, Xinglei Ma$^{*}$, Xuzhe Ying, K. T. Law$^{\dagger}$\\[.1cm]
	{\itshape ${}^1$Department of Physics, Hong Kong University of Science and Technology, Clear Water Bay, Hong Kong, China} 
\end{center}

\tableofcontents
\setcounter{equation}{0}
\setcounter{section}{0}
\setcounter{figure}{0}
\setcounter{table}{0}
\setcounter{page}{1}
\renewcommand{\theequation}{S\arabic{equation}}
\renewcommand{\thesection}{ \Roman{section}}
\renewcommand{\thefigure}{S\arabic{figure}}
\renewcommand{\thetable}{\arabic{table}}
\renewcommand{\tablename}{Supplementary Table}

\renewcommand{\bibnumfmt}[1]{[S#1]}
\renewcommand{\citenumfont}[1]{S#1}
\makeatletter

%%%%%%%%%%%%%%%%%%%%%%%%%%%%%%%%%%%%%%%%%%%%%%%%%%%%%%%%%%%%%%%%%%%%%%%%%%%%%%%%%%%%%%%%%%%%%%%%%%%%%%%%%%%%%%%%%%%%%%%%%%%%%%%%%%%%%%%%%%%%%%%%%%%%%%%%%%%%%%%%%%%%%%%%%%%%%%%%%%%%%%%%%%%%%%%%%%%%%%%%%%%%%%%%%%%%%%%%%%%%%%%%%%%%%%%%%%%%%%%%%%%%%%%%%%%%%%%%%%%%%%%%%%%%%%%%%%%%%%%%%%%%%%%%%%%%%%%%%%%%%%%%%%%%%%%%%%%%%%%%%%%%%%%%%%%%%%%%%%%%%%%%%%%%%%%%%%%%%%%%%%%%%%%%%%%%%%%%%%%%%%%%%%%%%%%%%%%%%%%%%%%%%%

\section*{\bf{\uppercase\expandafter{I. General relations between quantum metric and Majorana wavefunctions}}}
In this section, we show the general relations between the quantum metric of a set of isolated flat bands and the Majorana wavefunctions in a multiband system. Remarkably, our analysis reveals that the quadratic spread of the MZMs is determined by the quantum metric.
%=====================================================================================
\subsection*{A. Majorana wavefunctions by band projection}
To start, we consider a multiband Hamiltonian $H$, which consists of an unperturbed part $H_0$ and a perturbed part $H_1$:
\begin{equation}
	H=H_0+H_1
\end{equation}
with
\begin{equation}
	H_0 = \sum_{\bm{k}} \sum_{\alpha,\beta=1}^{N_h} h_{\alpha \beta} (\bm{k})  \hat{c}^\dagger_{\alpha}(\bm{k}) \hat{c}_{\beta}(\bm{k})
\end{equation}
and
\begin{equation}
	H_1 =  \frac{1}{\mathcal{N}} \sum_{\bm{r}\bm{r}^\prime} \sum_{\alpha,\beta=1}^{N_h}   U_{\alpha \beta}(\bm{r},\bm{r}^\prime) \hat{c}^\dagger_{\alpha}(\bm{r}) \hat{c}_{\beta}(\bm{r}^\prime)
	= \frac{1}{\mathcal{N}} \sum_{\bm{k}\bm{k}^\prime} \sum_{\alpha,\beta=1}^{N_h}   U_{\alpha \beta}(\bm{k},\bm{k}^\prime) \hat{c}^\dagger_{\alpha}(\bm{k}) \hat{c}_{\beta}(\bm{k}^\prime).
\end{equation}
Here, the $\hat{c}_{\alpha}(\bm{r}) $ is the annihilation operator of a fermion at position $\bm{r}$ with index $\alpha$. $\alpha=1,\cdots, N_h$ can denote the orbital  and particle-hole degrees of freedom. $N_h$ is the dimension of Hilbert space. $\mathcal{N}$ is the normalization factor. The term $U_{\alpha \beta}(\bm{r},\bm{r}^\prime)$ represents the perturbed Hamiltonian for creating the bound state, which could arise from factors such as a potential well or the boundary conditions. The Hamiltonian $H_0$ gives the band structure without any perturbation. If only a few isolated flat bands labeled by $n \in \mathcal{S} = \{1,2,\cdots N_f\}$ are close to the Fermi energy and they are significantly far away from other bands, we can project the fermionic operator $\hat{c}(\bm{k}) $ to the flat band manifold such that:
\begin{equation}
	\hat{c}_{\alpha}(\bm{k}) \rightarrow \sum_{n =1}^{N_f} g_{n,\alpha}(\bm{k}) \hat{a}_{n}(\bm{k}).
\end{equation}
Here $g_{n}(\bm{k}) = (g_{n,1}(\bm{k}),\cdots,g_{n,{N_h}}(\bm{k}))^T$ is an eigenvector of the Hamiltonian  $H_0$:
\begin{equation}
	\sum_{\beta} h_{\alpha \beta} (\bm{k})  g_{n,\beta}(\bm{k}) = \varepsilon_n(\bm{k}) g_{n,\alpha} (\bm{k}),
\end{equation}
which guarantees the diagonalized form of the $H_0$ after projection:
\begin{equation}
	H_0 \rightarrow \sum_{n=1}^{N_f} \sum_{\bm{k}} \varepsilon_n(\bm{k})\hat{a}^\dagger_n(\bm{k}) \hat{a}_{n}(\bm{k}). 
\end{equation}
Accordingly, we can do projection for the perturbed part:
\begin{equation}
	H_1 \rightarrow \frac{1}{\mathcal{N}}
	\sum_{n,n^\prime=1}^{N_f} \sum_{\bm{k},\bm{k}^\prime} \sum_{\alpha,\beta=1}^{N_h}  g_{n,\alpha}^*(\bm{k}) U_{\alpha \beta}(\bm{k},\bm{k}^\prime) g_{n^\prime,\beta}(\bm{k}^\prime) \hat{a}^\dagger_n(\bm{k}) \hat{a}_{n^\prime}(\bm{k}^\prime) .
\end{equation}
The total Hamiltonian after projection can be written as:
\begin{equation}\label{eq_bs_H}
	H = \frac{1}{\mathcal{N}} \sum_{n,n^\prime=1}^{N_f} \sum_{\bm{k},\bm{k}^\prime} \left(\varepsilon_{n}(\bm{k})  \delta_{nn^\prime} \delta_{\bm{k}\bm{k}^\prime}+ \sum_{\alpha,\beta=1}^{N_h} g_{n,\alpha}^*(\bm{k}) U_{\alpha\beta}(\bm{k},\bm{k}^\prime) g_{n^\prime,\beta}(\bm{k}^\prime)  \right) \hat{a}^\dagger_n(\bm{k}) \hat{a}_{n^\prime}(\bm{k}^\prime) .
\end{equation}

If the energy scale induced by the perturbation is much smaller than the band gap between the dispersive bands and the isolated flat bands, one can express the bound state wavefunction $\psi(\bm{r})$ as a linear combination of the isolated flat band Bloch wavefunctions $\psi_{n\bm{k}}(\bm{r})$ with the coefficients $w_{n\bm{k}}$:
\begin{equation}\label{eq_bs_wf}
	\psi_\alpha(\bm{r}) = \frac{1}{\mathcal{N}} \sum_{n=1}^{N_f} \sum_{\bm{k}} w_{n\bm{k}} \psi_{n\bm{k}}(\bm{r})
	= \frac{1}{\mathcal{N}}  \sum_{n=1}^{N_f}\sum_{\bm{k}} w_{n\bm{k}} g_{n,\alpha}(\bm{k}) e^{i\bm{k}\bm{r}} .
\end{equation}
In the last step, we expand the Bloch wavefunctions with atomic orbitals by considering $\phi_{\alpha}(\bm{r}-\bm{R}_i) = \delta(\bm{r}-\bm{R}_i)$. The bound state with energy $E$ satisfy the Schr\"odinger equation $H\psi(\bm{r}) = E\psi(\bm{r})$. By substituting Eq.~\eqref{eq_bs_H} and Eq.~\eqref{eq_bs_wf} to the Schr\"odinger equation and simplifying it, we find the coefficients satisfy: 
\begin{equation}
	\frac{1}{\mathcal{N}}\sum_{n\prime=1}^{N_f}\sum_{\bm{k}^\prime} \left(\varepsilon_{n}(\bm{k})  \delta_{nn^\prime} \delta_{\bm{k}\bm{k}^\prime} + \sum_{\alpha,\beta} g_{n,\alpha}^*(\bm{k}) U_{\alpha\beta}(\bm{k},\bm{k}^\prime) g_{n^\prime,\beta}(\bm{k}^\prime)  \right) w_{n^\prime \bm{k}^\prime} =  E w_{n \bm{k}} .
\end{equation}
Finally, the general form  of bound state wavefunction with multiband components is
\begin{equation}
	\begin{aligned}
		\psi_\alpha(\bm{r}) = \frac{1}{\mathcal{N}}\sum_{n=1}^{N_f}\sum_{\bm{k}} w_{n\bm{k}} g_{n,\alpha}(\bm{k}) e^{i\bm{k}\bm{r}}
		= \frac{1}{\mathcal{N}^2}\sum_{n=1}^{N_f}\sum_{\bm{k}} \sum_{\beta,\gamma=1}^{N_h} \frac{g_{n,\alpha}(\bm{k}) g_{n,\beta}^*(\bm{k}) e^{i\bm{k}\bm{r}}}{E-\varepsilon_{n}(\bm{k})} \sum_{n\prime=1}^{N_f} \sum_{\bm{k}^\prime}  U_{\beta\gamma}(\bm{k},\bm{k}^\prime) g_{n^\prime,\gamma}(\bm{k}^\prime) w_{n^\prime \bm{k}^\prime}  
	\end{aligned} .
\end{equation} 

Specifically, in the following, we will determine the Majorana wavefunctions of a one dimensional (1D) multiband topological superconductor (TSC) with an open boundary condition. To solve the open boundary problem, we use the Green's function method as shown in Ref.\cite{s_PhysRevB.83.125109}.  The Hamiltonian $H_0$ is the BdG Hamiltonian in Nambu basis $\hat{\Psi}(k)$.  There are totally $N_f$ flat Bogoliubov quasiparticle bands close to the Fermi energy and these bands are separated from other quasiparticle bands by some band gaps. MZMs are located at zero energy, and their wavefunctions can be constructed by the Bloch wavefunctions of these isolated flat quasiparticle bands. The perturbed Hamiltonian $H_1$ is set to cancel the coupling between the first unit cell ($1$) and the last unit cell ($N$) $U_{01}^{(\dagger)}$ to create an open boundary condition:
\begin{equation}
	H_1 = -\hat{\Psi}^\dagger(a) U_{01}^\dagger\hat{\Psi}(Na) - \hat{\Psi}^\dagger(Na) U_{01} \hat{\Psi}(a) 
	= \frac{1}{N} \sum_{kk^\prime}  \hat{\Psi}^\dagger(k) U(k,k^\prime) \hat{\Psi}(k^\prime).
\end{equation} 
Here, $U(k,k^\prime) = -U_{01}^\dagger e^{-ika}-U_{01}e^{ik^\prime a}$ is the momentum space representation of the perturbation $H_1$. $N$ is the number of sites of the 1D topological superconductor. The Majorana wavefunctions can be self-consistently expressed as:
\begin{equation}\label{eq_MZM_wavefunc}
	\begin{aligned}
		\psi(x) 
		& = -\frac{1}{N} \sum_{n=1}^{N_f} \sum_{k} \frac{g_{n}(k) g^\dagger_{n}(k)}{E-\varepsilon_{n}(k)} e^{ik(x-Na)} U_{01}\psi(a) -\frac{1}{N} \sum_{n=1}^{N_f}\sum_{k} \frac{g_{n}(k) g^\dagger_{n}(k)}{E-\varepsilon_{n}(k)}e^{ik(x-a)} U_{01}^\dagger \psi(Na)\\
		& = -G^p(x,Na;E) U_{01}\psi(a) -G^p(x,a;E) U_{01}^\dagger \psi(Na),
	\end{aligned}
\end{equation}
where $E=0$ for the MZMs and 
\begin{equation}
	G^p(x_j,x_{j^\prime};E) = \frac{1}{N} \sum_{n=1}^{N_f} \sum_{k} \frac{g_{n}(k) g^\dagger_{n}(k)}{E-\varepsilon_{n}(k)} e^{ik(x_j-x_{j^\prime})}
	\label{eq_projected_GF}
\end{equation}
denotes the projected Green function. The projected Green function $ G^p(x_j,x_{j^\prime};E)$ encodes the quantum metric effect induced by $g_{n}(k) g^\dagger_{n}(k)$, whose poles give additional decaying modes to the Majorana wavefunctions in addition to the poles of $1/(E-\varepsilon_n(k))$. %To solve the Majorana wavefunctions self-consistently, firstly, we set $x=a$ and $x=Na$ in Eq.~(\ref{eq_MZM_wavefunc}) and get two sets of solutions of $\psi(a)$ and $\psi(Na)$. Because these two energy states are degenerate and we can construct self-Hermitian Majorana wavefunctions out of $\psi(a)$ and $\psi(Na)$. Substituting these two sets of boundary wavefunctions to Eq.~(\ref{eq_MZM_wavefunc}) results in two Majorana (or fermionic) wavefunctions. We will focus on a single fermionic mode in the following text, because the other is simply its particle-hole partner. 

%=====================================================================================
%=====================================================================================
\subsection*{B. General relation between quadratic spread of MZMs and quantum metric}
\label{subsection_B}
To further study the relation between quantum metric and Majorana wavefunctions, we calculate the quadratic spread $\mathcal{QS}$ for each MZM. For the calculation of $\mathcal{QS}$, we denote the two MZM wavefunctions in Eq.~(\ref{eq_MZM_wavefunc}) as
\begin{equation}
	\psi_{\text{L}}(x) \equiv  -G^p(x,Na;E) U_{01}\psi(a) 
	\quad \text{and} \quad 
	\psi_{\text{R}}(x) \equiv  -G^p(x,a;E) U_{01}^\dagger \psi(Na).
\end{equation}
up to a phase factor. Now we take the left MZM as an example and show that its quadratic spread is determined by the quantum metric tensor. We define the quadratic spread for the left mode $\mathcal{QS}_{\text{L}}$ with respect to the left boundary site $x_{\text{L}}$:
\begin{equation}
	\begin{split}
		\mathcal{QS}_{\text{L}} &= \sum_{x=a}^{Na} (x-x_{\text{L}})^2 |\psi_{\text{L}}(x)|^2 \\
		&= \braket{x^2}_{\text{L}} -x_{\text{L}}'^2
	\end{split}
\end{equation}
where $\braket{x^2}_{\text{L}} = \sum_{x=a}^{Na} x^2 |\psi_{\text{L}}(x)|^2 $. $x_{\text{L}}'$ differs from $x_{\text{L}}$ by a model-dependent constant in the order of lattice constant $a$. For instance, in the Lieb-Kitaev model, this constant becomes zero in the flat quasiparticle band limit as shown in section\ref{subsection_QS}. So, we will neglect it in the following discussion.

For MZMs ($E=0$) in cases of completely flat quasiparticle bands ($\varepsilon_{n}(k)=\varepsilon_{n}$), 
\begin{equation}
	\begin{aligned}  
		\braket{x^2}_{\text{L}} 
		&=  \sum_{x=a}^{Na}  x^2 |\psi_{\text{L}}(x)|^2  \\
		&= \frac{1}{N^2} \sum_{x=a}^{Na}  \sum_{n,n'}^{N_f}  \sum_{k,k'} \frac{x^2 e^{i(k-k')(x-Na)}}{(E-\varepsilon_{n}(k))(E-\varepsilon_{n'}(k'))} \operatorname{Tr} \left[ g_{n}(k) g^\dagger_{n}(k) (-U_{01}) \psi(a) \psi^{\dagger}(a) (-U_{01}^\dagger) g_{n'}(k') g^\dagger_{n'}(k') \right] \\
		&= \frac{1}{N^2} \sum_{n,n'}^{N_f} \frac{U^2_{max}}{\varepsilon_{n}\varepsilon_{n'}} |\psi_A(a)|^2  \sum_{x=a}^{Na} \sum_{k,k'} x^2 e^{i(k-k')x}  \left[\braket{g_{n'}(k') | g_{n}(k)} \braket{g_{n}(k) | t_U | g_{n'}(k')}  \right].
	\end{aligned}
	\label{Eq_x2_MZM}
\end{equation}

Notice that in the last step, we extract the largest relevant energy scale $U_{max}^2$ and local density of states (LDOS) of one orbital (denoted as $A$) $|\psi_A(a)|^2$ by defining $U_{01} \psi(a) \psi^{\dagger}(a) U_{01}^\dagger \equiv U^2_{max}t_U  |\psi_A(a)|^2$ to make $t_U$ dimensionless.  It is worth noting that the prefactor $U^2_{max}/{\varepsilon}^2_{n}$, originated from MZMs' wavefunctions in Eq.~(\ref{eq_MZM_wavefunc}), provides the validity of the projection. To be specific, the construction of wavefunctions of MZMs by isolated flat bands is available when $U^2_{max}/\varepsilon^2_{n} \ll 1$ for any other band $n \notin \mathcal{S}$, which will be more clear in the Lieb-Kitaev model.

In general,  $t_U$ is a rank-1 matrix with model-dependent entries. However, as we will prove in section\ref{subsection_QS}, $t_U$ functions the same as a constant number $c_0$ in the context of $\braket{g_{n}(k) | t_U | g_{n'}(k')}$. That is,
\begin{equation}
	\braket{g_{n}(k) | t_U | g_{n'}(k')} \backsimeq c_0\braket{g_{n}(k) | g_{n'}(k')}.
	\label{Eq_tU_identity_approx}
\end{equation}
With this approximation,
\begin{equation}
	\begin{aligned}
		& \quad \frac{1}{N^2}\sum_{x=a}^{Na} \sum_{k,k'} x^2 e^{i(k-k')x} \braket{g_{n'}(k') | g_{n}(k)} \braket{g_{n}(k) |t_U| g_{n'}(k')} \\
		&\backsimeq \frac{c_0}{N^2} \sum_{x=a}^{Na} \sum_{k,k'} x^2 e^{i(k-k')x} \braket{g_{n'}(k') | g_{n}(k)} \braket{g_{n}(k) | g_{n'}(k')}\\
		&= \frac{c_0}{N^2} \sum_{x=a}^{Na} \sum_{k,k'} e^{i(k-k')x}
		\partial_k \partial_{k'} \left[\braket{g_{n'}(k') | g_{n}(k)}
		\braket{g_{n}(k) | g_{n'}(k')}\right]\\
		&= \frac{2c_0}{N} \sum_{k} \left[ 
		\braket{\partial_{k} g_{n'}(k) | \partial_k g_{n}(k)} \delta_{nn'}
		- \braket{\partial_{k} g_{n'}(k) |  g_{n}(k)} \braket{ g_{n}(k) |\partial_k g_{n'}(k)} \right] \\
		&= \frac{2c_0}{N}  \sum_k \mathcal{G}^{n n'}(k).
		\label{Eq_x_square_QM}
	\end{aligned}
\end{equation}
Here, $\mathcal{G}^{n n'}(k) = \braket{\partial_{k} g_{n'}(k) | \partial_k g_{n}(k)} \delta_{nn'}  - \braket{\partial_{k} g_{n'}(k) | g_{n}(k)} \braket{g_{n}(k) |\partial_k g_{n'}(k)}$ is the quantum geometric tensor, which is real and equivalent to quantum metric in one dimension. For $n=n'$, we recover the usual intra-band quantum metric.

Finally, the quadratic spread for the left MZM $\mathcal{QS}_{\text{L}}$ is
\begin{equation}
	\mathcal{QS}_{\text{L}} 
	\backsimeq  2c_0\sum_{n,n'}^{N_f} \frac{U^2_{\text{max}}}{\varepsilon_{n}\varepsilon_{n'}} |\psi_A(a)|^2   \frac{1}{N} \sum_k \mathcal{G}^{n n'}(k). % - x_\text{L}'^2.
	\label{QS_L}
\end{equation}

Similar relation can be obtained for the right MZM:
\begin{equation}
	\mathcal{QS}_{\text{R}} 
	\backsimeq  2c_0\sum_{n,n'}^{N_f} \frac{U^2_{\text{max}}}{\varepsilon_{n}\varepsilon_{n'}} |\psi_A(Na)|^2   \frac{1}{N} \sum_k \mathcal{G}^{n n'}(k). % - x_\text{R}'^2.
	\label{QS_R}
\end{equation}
The above general relations Eq.(\ref{QS_L})-Eq.(\ref{QS_R}) imply that the spread of Majorana zero mode is essentially controlled by the quantum metric of quasiparticle bands. We will show this relation in detail in the Lieb-Kitaev model but with quantum metric being the normal bands'.

%%%%%%%%%%%%%%%%%%%%%%%%%%%%%%%%%%%%%%%%%%%%%%%%%%%%%%%%%%%%%%%%%%%%%%%%%%%%%%%%%%%%%%%%%%%%%%%%%%%%%%%%%%%%%%%%%%%%%%%%%%%%%%%%%%%%%%%%%%%%%%%%%%%%%%%%%%%%%%%%%%%%%%%%%%%%%%%%%%%%%%%%%%%%%%%%%%%%%%%%%%%%%%%%%%%%%%%%%%%%%%%%%%%%%%%%%%%%%%%%%%%%%%%%%%%%%%%%%%%%%%%%%%%%%%%%%%%%%%%%%%%%%%%%%%%%%%%%%%%%%%%%%%%%%%%%%%%%%%%%%%%%%%%%%%%%%%%%%%%%%%%%%%%%%%%%%%%%%%%%%%%%%%%%%%%%%%%%%%%%%%%%%%%%%%%%%%%%%%%%%%%%%%
\section*{\bf{\uppercase\expandafter{II. MZMs in the Lieb-Kitaev Model}}}
In this section, we obtain Majorana wavefunctions of the Lieb-Kitaev model in the isolated flat bands limit. We point out that the effective two-band Hamiltonian is exactly the same as that of the single band Kitaev model, however, in a completely different basis. Furthermore, we analytically show the proportionality between quadratic spread of MZMs and the quantum metric length. We simplify the projected Green's function and find that the localization length of the MZMs is proportional to the quantum metric length.
%================================================================================================================================================
\subsection*{C. The Lieb-Kitaev model}\label{subsection_C}
To be specific, we consider the Lieb-Kitaev Hamiltonian.  The total Hamiltonian $H$ includes the original unperturbed Hamiltonian $H_0$ and the boundary condition perturbation $H_1$. The unperturbed Hamiltonian is exactly the BdG Hamiltonian shown in the main text:
\begin{equation} \label{Eq_BdG_majorana}
	H_0 = H_{\text{BdG}} = \sum_{k} \hat{\Psi}^\dagger(k) \mathcal{H}(k)\hat{\Psi}(k) = \sum_{k} \hat{\Psi}^\dagger(k) \begin{pmatrix}h(k)& -i2\Delta \sin{(ka)} \mathbb{I}_3\\
		i2\Delta \sin{(ka)} \mathbb{I}_3 &  -h^*(-k)\end{pmatrix}\hat{\Psi}(k),
\end{equation}
with \begin{equation} 
	h(k) =  (-\mu-2t\cos{(ka)})\mathbb{I}_3 +\begin{pmatrix}
		-V& a_k & 0 \\
		a_k^*& 0& a_k^* \\
		0& a_k & V 
	\end{pmatrix} 
	\quad \text{and} \quad a_k = -J(1+ e^{ika}).
\end{equation}
The perturbed Hamiltonian $H_1$ that gives open boundary conditions is 
\begin{equation}
	H_1  = \frac{1}{\mathcal{N}}\sum_{kk^\prime}  \hat{\Psi}^\dagger(k) U(k,k^\prime) \hat{\Psi}(k^\prime).
\end{equation} 
Here, $U(k,k^\prime) = 
-U_{01}^\dagger e^{-ika}-U_{01}e^{ik^\prime a}$. $U_{01}$ represents the coupling between two ends:
\begin{equation}
	U_{01} = \begin{pmatrix} h_{01} & -\tilde{\Delta}\\ \tilde{\Delta} &-h_{01}^* \end{pmatrix}
	\rightarrow \tau_z \otimes h_{01} - i\tau_y \otimes \tilde{\Delta}, 
\end{equation}
by assuming the positive and real hoping and pairing for simplicity, with 
\begin{equation}
	h_{01} = \begin{pmatrix}
		-t& -J & 0 \\ 0& -t& 0 \\ 0& -J &  -t \end{pmatrix} \quad \text{and} \quad
	\tilde{\Delta}=\begin{pmatrix}
		\Delta & 0 & 0 \\ 0& \Delta& 0 \\ 0& 0 &  \Delta \end{pmatrix}.
	\label{Eq_h_01}
\end{equation}
$\tau_i$ is Pauli matrix. $H_1$ removes the hopping and pairing between the first site $1$ and the last site $N$ from $H_0$.  

When $V\gg t \simeq \Delta$, we have two isolated flat quasiparticle bands labeled by $n=\pm$, forming our flat band manifolds which we will project onto. The Majorana wavefunctions can be constructed by the Bloch eigenvectors $g_{\pm}(k)$ of $n=\pm$ bands with dispersion $\varepsilon_\pm(k)= \pm \sqrt{\left(-\mu-2t\cos{(ka)}\right)^2 +4\Delta^2\sin^2{(ka)}}$. In the following subsections, we will work on this Lieb-Kitaev model, and study its topological property as well as quantum metric effect.
%================================================================================================================================================
\subsection*{D. Topological invariant of the Lieb-Kitaev model}
Topological phases can be characterized by topological invariants. According to the 10-fold topological classification \cite{s_AYuKitaev_2001,s_PhysRevB.78.195125}, the one dimensional single band Kitaev model is characterized by the Z2 topological invariant belonging to the class D. The Lieb-Kitaev model belongs to the same class and can also be distinguished by the Z2 number $\mathcal{Q}$ that is defined by the Pfaffian of the BdG Hamiltonian in Eq.~(\ref{Eq_BdG_majorana}):
\begin{equation}
	\mathcal{Q} = \operatorname{sign}(\text{Pf}[i\tilde{H}(k=0)]\text{Pf}[i\tilde{H}(k=\frac{\pi}{a})]).
	\label{Eq_topo_number}
\end{equation}
The topological invariant only involves the two states with momentum $k=0$ and $k=\pi/a$ by the fact that only these two states can switch the \textit{fermion parity} of the ground state, a quantity which directly determines the topological property. To obtain the Pfaffian of the BdG Hamiltonian, we perform a unitary transformation on the momentum space BdG Hamiltonian $\mathcal{H}(k)$ in Eq.~(\ref{Eq_BdG_majorana}) to obtain an antisymmetric one $\tilde{H}(k)$. Specifically, there are two steps for the unitary transformation: 
\begin{equation}
	\tilde{H}(k) = U^a X(k) \mathcal{H}(k) X^\dagger(k) {U^a}^{\dagger},
\end{equation}
Firstly, we use $X$ to transform $\mathcal{H}(k)$ from orbital to normal band basis. $X$ is the unitary matrix consisting of normal band eigenvectors:
\begin{equation}
	X(k) = \begin{pmatrix}
		& u_-(k) & 0 &u_0(k) & 0 &u_+(k)&0\\
		&0 & u_-^*(-k) & 0 &u_0^*(-k) & 0 &u_+^*(-k)\\
	\end{pmatrix},
\end{equation}
with $h(k) u_n(k) = \epsilon_n(k) u_n(k)$. Then, we further antisymmetrize the matrix by $U^a$:
\begin{equation}
	U^a = \frac{1}{\sqrt{2}} \begin{pmatrix}  1 & 1 & & & &  \\  i & -i & & & &   \\
		&  & 1 & 1 & & \\   &  & i & -i & & \\
		& & & & 1 & 1  \\   & &  &  & i & -i \\
	\end{pmatrix}.
\end{equation}
With the transformation, one can write down $\tilde{H}(k=0)$ and $\tilde{H}(k=\frac{\pi}{a})$ as:
\begin{equation}
	\begin{split}
		\tilde{H}(k=0) &= -i \text{Bdiag} \begin{pmatrix}  
			& \begin{pmatrix}  
				0 & -\mu-2t-\sqrt{V^2+8J^2}  \\  \mu+2t+\sqrt{V^2+8J^2} & 0   
			\end{pmatrix}, \\
			& \begin{pmatrix}
				0 & -\mu-2t  \\ \mu+2t & 0 
			\end{pmatrix}, \\
			& \begin{pmatrix}
				0 & -\mu-2t+\sqrt{V^2+8J^2}   \\   \mu+2t-\sqrt{V^2+8J^2} & 0 \\
			\end{pmatrix}
		\end{pmatrix}
		\\
		\tilde{H}(k=\frac{\pi}{a}) &= -i \text{Bdiag} \begin{pmatrix}  
			& \begin{pmatrix}  
				0 & -\mu+2t-V  \\  \mu-2t+V & 0   
			\end{pmatrix}, \\
			& \begin{pmatrix}
				0 & -\mu+2t  \\ \mu-2t & 0 
			\end{pmatrix}, \\
			& \begin{pmatrix}
				0 & -\mu+2t+V  \\ \mu-2t-V & 0 \\
			\end{pmatrix}
		\end{pmatrix}.
	\end{split}
\end{equation}
Then we can easily obtain:
\begin{equation}
	\begin{split}
		\text{Pf}[i\tilde{H}(k=0)] &= (-\mu-2t-\sqrt{V^2+8J^2})(-\mu-2t)(-\mu-2t+\sqrt{V^2+8J^2})) \\
		\text{Pf}[i\tilde{H}(k =\frac{\pi}{a})] &= (-\mu+2t-V)(-\mu+2t)(-\mu+2t+V)).
	\end{split}
\end{equation}
Given the definition of the Z2 topological number in Eq.~(\ref{Eq_topo_number}), we find that the Lieb-Kitaev model is topological nontrivial under either the following condition with $\mathcal{Q} = -1$: 
\begin{equation}
	\begin{aligned}	
		-\sqrt{V^2+8J^2}-2t < &\mu < -V+2t,\\ 
		-2t<&\mu <2t, \\
		V+2t < &\mu < \sqrt{V^2+8J^2}-2t . 
	\end{aligned}
	\label{Eq_topo_condition}
\end{equation}
Namely, whenever the chemical potential cuts through odd number of the normal bands, this model enters the topological phase and holds MZMs under open boundary condition. 

One can figure out the relation between the topological invariant and the existence of zero energy modes from following argument. As the Pfaffian $\text{Pf}[i\tilde{H}(k)]$ determines the fermion parity of the Hamiltonian matrix $\mathcal{H}(k)$, the nontrivial topological number $\mathcal{Q} = -1$ means that states $k=0$ and $k=\pi/a$ have distinct fermion parities. When continuously tuning the $\mathcal{H}(0)$ into $\mathcal{H}(\pi/a)$, one has to encounter a zero-energy state. This fact implies the existence of zero-energy states under open boundary condition (OBC). A practical way to see this is to continuously tune the boundary condition from periodic (PBC) to anti-periodic (APBC), which allows it to go through a point with exact OBC. The crystal momentum for PBC and APBC are $k=2\pi n/Na$ and $k=\pi(2n+1)/Na $ with $n \in \mathbb{Z}$ and $N$ is the site number. Notice that $k=0$ is only present in PBC, and $k=\pi/a$ exists in PBC (APBC) if there are even (odd) number of sites, thus the ground state fermion parities of the system under PBC and APBC will be different for nontrivial $\mathcal{Q}=-1$ regardless of the site number parity. This means that at the middle point of the tuning when OBC is created, there will be unpaired zero-energy Majorana edge modes. For more discussion about Pfaffian and MZMs, one can refer to \cite{s_AYuKitaev_2001}.

In addition, as shown in Eq.~(\ref{Eq_topo_condition}), the topological conditions work individually for each normal band, which implies that the low-energy effective model  involving only the projected nearly flat band would capture the essential physics well. This will be done in the later subsection.

%================================================================================================================================================
\subsection*{E. Quantum metric length of the Lieb-Kitaev model}
While the topological property determines the existence of MZMs, the quantum metric controls the MZMs' spatial behavior. Specifically, it provides an important length scale, called as quantum metric length (QML), which determines the localization length and quadratic spread of the MZMs. In this subsection we analytically calculate QML, which will show up frequently in the later discussion of quadratic spread and localization length.

The Brillouin Zone averaged quantum metric of the flat band in normal Hamiltonian $h(k)$ can be evaluated analytically. For 1D Lieb-like model, with the eigenvector of the nearly flat band of normal Hamiltonian $h(k)$:
\begin{equation}\label{equ0}
	u_0(k) = \frac{1}{\sqrt{4J^2(1+\cos{(ka)})+V^2}} \begin{pmatrix}
		a_k \\V\\-a_k
	\end{pmatrix},
\end{equation}
we can calculate quantum metric $\mathcal{G}^0(k)$ and Brillouin Zone averaged quantum metric $\bar{\mathcal{G}}$ analytically:
\begin{equation}
	\mathcal{G}^0(k)  = \bra{\partial_{k} u_0 (k)}\left[1-\ket{u_0 (k)}\bra{u_0 (k)} \right]\ket{\partial_k u_0 (k)} = \frac{2V^2/ J^2 }{\left(4(1+\cos{(ka)})+V^2/J^2\right)^2} a^2
	\label{Eq_Quantum_Metric}
\end{equation}
and
\begin{equation}
	\bar{\mathcal{G}} = \frac{a}{2\pi} \int dk \mathcal{G}^0(k)
	= \frac{\sqrt{2}}{4}\frac{1+\frac{V^2}{4J^2}}{\frac{V}{J} \left( \frac{V^2}{8J^2}+1\right)^{3/2}} a^2.
\end{equation}
QML is defined as
\begin{equation}    \label{eq_qmlength}
	\xi_{\text{QM}} \equiv \frac{\bar{\mathcal{G}}}{a} 
\end{equation}
which goes to the limit $\xi_{\text{QM}}\rightarrow \frac{\sqrt{2}}{4} \frac{J}{V}$ when $\frac{V}{J}\rightarrow 0$.

%================================================================================================================================================
\subsection*{F. Low-energy two-band Hamiltonian}
In this subsection, we show that the low-energy two-band Hamiltonian of the Lieb-Kitaev model is the same as that of the single band Kitaev model, however, in a different basis. What's more, the Bloch eigenvectors of quasiparticle bands $g_{\pm}(k)$ can be expressed as the direct product of normal flat band Bloch eigenvector $u_{0}(k)$ and eigenvectors of Kitaev Hamiltonian $v_\pm(k)$, which helps to simplify the calculation of quadratic spread and localization length.

One can divide the projection of unperturbed Hamiltonian into two steps:
\begin{equation}
	\begin{aligned}
		H_0 & \rightarrow \sum_{k} \hat{\Psi}^\dagger(k)   W_1(k) 
		\begin{pmatrix} -\mu-2t\cos{(ka)} & -i2\Delta \sin{(ka)}  \\ i2\Delta \sin{(ka)} & \mu +2t\cos{(ka)} \end{pmatrix} W_1^\dagger(k) \hat{\Psi}(k) \\
		& = \sum_{k} \hat{\Psi}^\dagger(k)   W_1(k)  W_2(k) 
		\begin{pmatrix}  \varepsilon_-(k) & 0  \\ 0 & \varepsilon_+(k) \end{pmatrix} W_2^\dagger(k) W_1^\dagger(k) 
		\hat{\Psi}(k) \\
		& =  \sum_{k}\sum_{n=\pm}  \varepsilon_n (k) \hat{a}_n^\dagger(k) \hat{a}_n(k) .  
	\end{aligned} 
\end{equation}
In the first step, we project the $6\times 6$ Hamiltonian to two isolated quasiparticle bands by the $6\times2$ non-square matrix:
\begin{equation}\label{eqW1}
	W_1(k) = \begin{pmatrix} u_{0}(k) & 0\\ 0&u^*_{0}(-k)  \end{pmatrix},
\end{equation}
with $u_0(k)$ shown in Eq.~(\ref{equ0}). Specifically, $u_{0}(k)= u_{0}^*(-k)$ for the Lieb-Kitaev model. With this, the effective two band Hamiltonian is the same as that in the single band Kitaev model, in the basis $W_1^{\dagger}\hat{\Psi}(k)$:
\begin{equation}
	W_1^\dagger (k) \begin{pmatrix}  h(k) & -i2\Delta \sin{(ka)} \mathbb{I}_3\\ i2\Delta \sin{(ka)}  \mathbb{I}_3 & -h^*(-k) \end{pmatrix} W_1(k) = \left(- \mu-2t\cos{(ka)} \right) \tau_z + 2\Delta \sin{(ka)}\tau_y \equiv H_{\text{Kitaev}}(k). 
\end{equation}
In the second step, we diagonalize the Kitaev Hamiltonian by the matrix $W_2 = (v_-,v_+)$ with
$H_{\text{Kitaev}} v_\pm = \varepsilon_\pm(k) v_\pm$. Eigenvalues are $\varepsilon_\pm(k)= \pm \sqrt{\left(-\mu-2t\cos{(ka)}\right)^2 +4\Delta^2\sin^2{(ka)}}$, with eigenvectors:
\begin{equation}\label{eq_vpm}
	v_\pm(k) =  \frac{e^{i\phi_\pm(k)}}{\sqrt{(\varepsilon_\pm(k) +\mu+2 t\cos{(ka)})^2 +4\Delta^2\sin^2{(ka)}}} \begin{pmatrix}
		-i2\Delta \sin{(ka)} \\ \varepsilon_\pm(k)+\mu+ 2t\cos{ka}
	\end{pmatrix}
\end{equation}
$\phi_\pm(k)$ is an arbitrary phase. Therefore, the total projection matrix can be written as $ W=\left(g_-(k),g_+(k)\right)= W_1 W_2$.  Or equivalently, one can express the Bloch eigenvectors $g_{\pm}(k)$ in the direct product form:
\begin{equation} \label{eq_gvu}
	g_{\pm}(k) =  v_\pm(k) \otimes u_{0}(k).
\end{equation}
It is worth noting that although the low-energy two-band Hamiltonian of the Lieb-Kitaev model is the same as the single-band Kitaev model, quantum geometric contribution is kept in the wavefunctions.

%================================================================================================================================================
\subsection*{G. Quadratic spread of MZMs in the Lieb-Kitaev model}
\label{subsection_QS}
In this subsection, we apply the conclusion about quadratic spread in section\ref{subsection_B} to the Lieb-Kitaev model. We will show that the approximation used in Eq.~(\ref{Eq_tU_identity_approx}) is exactly right to the order of $\mathcal{O}(V/J)$ in the Lieb-Kitaev model. Then, we prove that the quadratic spread of the MZM is proportional to the QML as defined in Eq.~(\ref{eq_qmlength}). 

For the Lieb-Kitaev model, the projected MZMs' wavefunctions in Eq.~(\ref{eq_MZM_wavefunc}) can be further simplified by separating it to particle-hole part $\psi^v$ and orbital part $\psi^u$. Here, we take the right MZM $\psi_{\text{R}}(x)$ as an example:
\begin{equation}
	\begin{aligned}
		\psi_{\text{R}}(x) 
		&= -\frac{1}{N}\sum_{n=\pm} \sum_{k} \frac{e^{ik(x-a)}}{E-\varepsilon_{n}(k)}g_{n}(k) g_{n}^{\dagger}(k)  U_{01}^{\dagger} \psi(Na) \\
		& = 
		-\frac{1}{N} \sum_{n=\pm} \sum_{k} \frac{e^{ik(x-a)}}{E-\varepsilon_{n}(k)} \left\{\left[v_n(k) v_n^\dagger(k) \tau_z^{\dagger} \psi^{v}(Na) \right] \otimes \left[ u_{0}(k) u_{0}^\dagger(k) h_{01}^{\dagger} \psi^{u}(Na)\right] \right. \\
		& \ \ \ \ \ \quad\quad\quad + \left.\left[v_n(k) v_n^\dagger(k) (-i\tau_y)^{\dagger} \psi^{v}(Na) \right] \otimes \left[ u_{0}(k) u_{0}^\dagger(k) \tilde{\Delta}^{\dagger} \psi^u(Na)\right]     \right\} \\ 
		&= 
		\sum_{k} \overbrace{\left[ \frac{1}{\varepsilon_+(k)}\left(v_+(k)v_+^\dagger(k)-v_-(k)v_-^\dagger(k) \right) \tau_z^{\dagger} \psi^{v}(Na) \right]}^{\circled{1}} 
		\otimes \left[  e^{ik(x-a)}
		u_{0}(k) u_{0}^\dagger(k) \overbrace{h_{01}^{\dagger} \psi^{u}(Na)}^{\circled{2}}  \right]  \\
		& \ \ \ \ \ \quad\quad\quad + \sum_{k} \overbrace{\left[\frac{1}{\varepsilon_+(k)} \left( v_+(k)v_+^\dagger(k)-v_-(k)v_-^\dagger(k) \right)(-i\tau_y)^{\dagger} \psi^{v}(Na)     \right]}^{\circled{3}} 
		\otimes \left[ e^{ik(x-a)} u_{0}(k) u_{0}^\dagger(k) \overbrace{\tilde{\Delta}^{\dagger} \psi^{u}(Na) }^{\circled{4}} \right],
	\end{aligned}
\end{equation}
where we have used the fact that $E=0$ for MZMs and $\varepsilon_-(k)=-\varepsilon_+(k)$. $v_{\pm}$ and $u_0$ are given in Eq.~(\ref{eq_vpm}) and Eq.~(\ref{equ0}), respectively. $\psi(Na)$ can solved from Eq.~(\ref{eq_MZM_wavefunc}) and Eq.~(\ref{eq_projected_GF}) self-consistently. 

The particle-hole components of MZMs' wavefunctions at end site is $\psi^v(Na)=(1,-1)^{\operatorname{T}} $. Then, $\circled{1}$, $\circled{3}$ can be simplified as
\begin{equation} 
	\begin{aligned}
		&\circled{1}  
		= \left[  \frac{1}{\varepsilon_+^2(k)} \begin{pmatrix} -\mu-2t\cos{(ka)} & -i2 \Delta \sin{(ka)} \\  i2\Delta \sin{(ka)} & \mu+2t\cos{(ka)}  \end{pmatrix} 
		\begin{pmatrix} 1 & 0 \\  0 & -1  \end{pmatrix} 
		\begin{pmatrix} 1 \\  -1 \end{pmatrix}  \right] 
		= \frac{-\mu-2t\cos{(ka)} - i2 \Delta \sin{(ka)}}{\varepsilon_+^2(k)} \begin{pmatrix} 1 \\  -1  \end{pmatrix} \\
		&\circled{3}
		=\left[  \frac{1}{\varepsilon_+^2(k)} \begin{pmatrix} -\mu-2t\cos{(ka)} &-i2 \Delta \sin{(ka)} \\  i2\Delta \sin{(ka)} & \mu+2t\cos{(ka)}  \end{pmatrix} 
		\begin{pmatrix} 0 & 1 \\  -1 & 0  \end{pmatrix} 
		\begin{pmatrix} 1 \\  -1 \end{pmatrix}  \right] 
		= - \frac{-\mu-2t\cos{(ka)} - i2 \Delta \sin{(ka)}}{\varepsilon_+^2(k)} \begin{pmatrix} 1 \\  -1  \end{pmatrix} .
	\end{aligned}
\end{equation}

In the flat quasiparticle band limit with $t \backsimeq \Delta$, the orbital parts at two ends to the order of $\mathcal{O}(V/J)$ are
\begin{equation}
	\psi^u(Na) = \begin{pmatrix} 1 \\ V/J\\  -1 \end{pmatrix} \psi^u_A(Na).		
\end{equation}
Then, $\circled{2}$ and $\circled{4}$ becomes
\begin{equation}
	\begin{aligned}
		\circled{2} 
		&=   \begin{pmatrix}
			-t& 0 & 0 \\ -J& -t& -J \\ 0 & 0 &  -t \end{pmatrix}
		\begin{pmatrix} 1 \\ V/J \\  -1 \end{pmatrix}\psi^u_A(Na) 
		= -t \begin{pmatrix} 1 \\ V/J \\  -1  \end{pmatrix}\psi^u_A(Na) \\
		\circled{4} 
		& =  \begin{pmatrix}
			\Delta & 0 & 0 \\ 0& \Delta & 0 \\ 0& 0 &  \Delta
		\end{pmatrix}  
		\begin{pmatrix} 1 \\ V/J \\  -1 \end{pmatrix}\psi^u_A(Na) 
		= \Delta \begin{pmatrix} 1 \\ V/J \\  -1 \end{pmatrix}\psi^u_A(Na).
	\end{aligned}
	\label{psi_last}
\end{equation}

The right MZM wavefunction would be
\begin{equation}
	\begin{aligned}
		\psi_{\text{R}}(x) & 
		\xlongequal[t \backsimeq \Delta]{\mathcal{O}(V/J)}
		\sum_{k}
		\left[ \frac{-\mu-2t\cos{(ka)} - 2i \Delta \sin{(ka)}}{\varepsilon_+^2(k)} \begin{pmatrix} 1 \\  -1  \end{pmatrix}  \right] 
		\otimes  \left[  e^{ik(x-a)}
		u_{0}(k) u_{0}^\dagger(k) (-t-\Delta) \begin{pmatrix} 1 \\  V/J \\  -1 \end{pmatrix}\psi^u_A(Na) \right] \\   
		&\xlongequal[]{\mu=0,t=\Delta}
		\begin{pmatrix} 1 \\  -1  \end{pmatrix}
		\otimes \sum_{k} \left[  e^{ikx}
		u_{0}(k) u_{0}^\dagger(k) \begin{pmatrix} 1 \\ V/J \\  -1 \end{pmatrix}\psi^u_A(Na) \right].
	\end{aligned}
	\label{Eq_MZM_R_simplify}
\end{equation}

Here are some comments for the simplified MZM wavefunction in Eq.~(\ref{Eq_MZM_R_simplify}). Firstly, as shown in the first step of Eq.~(\ref{Eq_MZM_R_simplify}), $\max{(\Delta,t)}$ determines the survived largest energy scale in $U_{\text{max}}$ as introduced in section\ref{subsection_B}. Since it is much smaller than the dispersive band gap $V$, the validity of projection (or discarding higher bands) is guaranteed. Secondly, the particle-hole part in MZMs' wavefunctions is independent of momentum when quasiparticle bands $n=\pm$ are completely flat, which implies that the real space local density of states (LDOS) is only determined by the orbital part. In other words, as a special case of the general result in Eq.~(\ref{QS_R}), the quadratic spread is indeed directly determined by the normal band quantum metric as shown in the following.

Now we can proceed to calculate the quadratic spread of the right MZM. The LDOS of it is:
\begin{equation}\label{Eq_MZM_LDOS}
	\rho_{\text{R}}(x) =  2\braket{\psi_{\text{R}}(x)|\psi_{\text{R}}(x)} 
	= |\psi^u_A(Na)|^2 
	\sum_{k,k'} e^{i(k-k')x}
	\braket{u_{0}(k') |u_{0}(k)} 
	\braket{u_{0}(k) |t^{orb}_U|u_{0}(k')},  
\end{equation}
where $t_U$ to the order of $\mathcal{O}(V/J)$ is
\begin{equation}
	\begin{aligned}
		t^{orb}_U & =  \begin{pmatrix} 1 \\ V/J \\  -1 \end{pmatrix}  \begin{pmatrix} 1 & V/J &  -1 \end{pmatrix}   
		= 
		\begin{pmatrix} 1 & V/J & -1\\ V/J & 0 & -V/J \\  -1 & -V/J & 1 \end{pmatrix}.  
	\end{aligned}
\end{equation}
$t^{orb}_U$ is rank-1 matrix and functions as a constant number when considering the inner product $\braket{u_{0}(k) |t^{orb}_U|u_{0}(k')}$ to the order of $\mathcal{O}(V/J)$: 
\begin{equation}
	\begin{aligned}
		\braket{u_{0}(k) |t^{orb}_U|u_{0}(k')}  
		& =
		\frac{1}{\sqrt{2|a_k|^2+V^2}\sqrt{2|a_{k'}|^2+V^2}}\begin{pmatrix} a_k^* & V &  -a_k^* \end{pmatrix}
		\begin{pmatrix} 1 & V/J & -1\\ V/J & 0 & -V/J \\  -1 & -V/J & 1 \end{pmatrix}
		\begin{pmatrix} a_{k'} \\ V \\  -a_{k'} \end{pmatrix}\\ 
		&=\frac{4 a_k^* a_{k'}}{\sqrt{2|a_k|^2+V^2}\sqrt{2|a_{k'}|^2+V^2}} \\
		&=   2  \frac{2 a_k^* a_{k'} + V^2}{\sqrt{2|a_k|^2+V^2}\sqrt{2|a_{k'}|^2+V^2}} + \mathcal{O}\left((V/J)^2\right) \\
		&=   2  \braket{u_{0}(k) | u_{0}(k')} + \mathcal{O} \left((V/J)^2\right).
	\end{aligned}
\end{equation}
Therefore, to order $ \mathcal{O}(V/J)$, the LDOS for the right MZM is simplified as
\begin{equation} 
	\rho_{\text{R}}(x) = 4 |\psi^u_A(Na)|^2 
	\sum_{k,k'} e^{i(k-k')x}
	|\braket{u_{0}(k') |u_{0}(k)}|^2. 
\end{equation}

Applying the proof in Eq.~(\ref{Eq_x_square_QM}), we obtain the relation between the quadratic spread and the normal band quantum metric:
\begin{equation}
	\begin{aligned}
		\mathcal{QS}_{\text{R}} 
		&= \sum_{x=a}^{Na} (x-Na)^2 \rho_{1}(x)  \\
		& = 4  |\psi^u_A(Na)|^2 \sum_{x=a}^{Na} (x-Na)^2     \sum_{k,k'} e^{i(k-k')x}
		\left|\braket{u_{0}(k') |u_{0}(k)}\right|^2   \\
		&= 8  |\psi^u_A(Na)|^2   \sum_k \mathcal{G}(k) \\
		&= 8  |\psi^u_A(Na)|^2   a  \xi_{\text{QM}}. 
	\end{aligned}  
\end{equation}
where we have used the definition for QML in Eq.~(\ref{eq_qmlength}). With the normalization constant obtained from Eq.~(\ref{eq_MZM_wavefunc}) and Eq.~(\ref{eq_projected_GF}) self-consistently, $\psi^u_A(Na) = 0.3536 \approx 1/2\sqrt{2}$, which is independent of $V/J$. Finally, we find
\begin{equation}
	\mathcal{QS}_{\text{R}} =  a  \xi_{\text{QM}},
\end{equation} 
which matches perfectly with numerical result by diagonalizing the real-space lattice model shown in the main text.

%================================================================================================================================================
\subsection*{H. Localization length of MZMs in the Lieb-Kitaev model}
In this subsection, we simplify the projected Green function matrix $G^p(x_j,x_{j^\prime};E=0)$ and find out the exact form of Majorana wavefunctions and localization lengths. Besides the traditional localization length $\xi_{\text{BCS}} = -a/\ln|\sqrt{\frac{t-\Delta}{t+\Delta}}|$ at $\mu=0$, an additional localization length $\xi_+=4\xi_{\text{QM}}$ emerges.

By expressing the quasiparticle bands Bloch eigenvectors $g_{\pm}(k)$ as the direct product form in Eq.~(\ref{eq_gvu}), we can simplify the projected Green function $G^p(x_j,x_{j^\prime};E=0)$ dramatically: 
\begin{equation}
	\begin{aligned}
		G^p(x_j,x_{j^\prime};E=0)
		& = -\frac{1}{N}\sum_{k} \frac{v_+(k)v_+^\dagger(k)-v_-(k)v_-^\dagger(k)}{\varepsilon_+(k)} \otimes u_{0}(k)u_{0}^\dagger(k)  e^{ika(j-j^\prime)} \\
		& =- \frac{1}{N} \sum_{k} \frac{1}{\varepsilon_+(k)^2} \begin{pmatrix} -\mu-2t\cos{(ka)} &-i2 \Delta \sin{(ka)} \\  i2\Delta \sin{(ka)} & \mu+2t\cos{(ka)}  \end{pmatrix} \otimes u_{0}(k)u_{0}^\dagger(k) e^{ika(j-j^\prime)} \\
		& = -\frac{1}{N} \sum_{k} \frac{1}{\left(\mu+2t\cos{ka}\right)^2 + 4\Delta^2\sin^2{(ka)}} \begin{pmatrix} -\mu-2t\cos{(ka)} &-i2 \Delta \sin{(ka)} \\  i2\Delta \sin{(ka)} & \mu+2t\cos{(ka)}  \end{pmatrix} \otimes\\  &\quad\quad\quad 	\frac{1}{4(1+\cos{(ka)})+\frac{V^2}{J^2}} \begin{pmatrix}
			2(1+\cos{(ka)})  & -\frac{V}{J}(1+e^{ika}) &-2(1+\cos{(ka)})
			\\	h.c.  &\frac{V^2}{J^2}  & \frac{V}{J}(1+e^{-ika}) 
			\\  h.c.&h.c.& 2(1+\cos{(ka)})  \end{pmatrix} e^{ika(j-j^\prime)}.
	\end{aligned} 
\end{equation}
For $j>j^\prime$, we substitute $z=e^{ika}$ and do the contour integral:
\begin{equation}\label{Gpgeq}
	\begin{aligned}
		G^p(x_j\geq x_{j^\prime};E=0)
		& = \frac{1}{2} \frac{1}{\Delta^2-t^2} \ointctrclockwise_{|z|=1} \frac{dz}{2\pi i } \frac{z^{j-j^\prime}}{(z -z_1) (z -z_2) (z -z_3)(z -z_4)  (z-z_+)(z-z_-)}
		\\  &  \quad\quad\quad \times \begin{pmatrix} -\mu z-t(z^2+1)  &-\Delta (z^2-1) \\  \Delta (z^2-1) & \mu z+ t (z^2+1)  \end{pmatrix} \otimes
		\begin{pmatrix} (z+1)^2  & -\frac{V}{J}z(1+z) & -(z+1)^2 
			\\-\frac{V}{J}(z+1)  &\frac{V^2}{J^2}  & \frac{V}{J}(z+1) 
			\\ -(z+1)^2 & \frac{V}{J}z(1+z) & (z+1)^2
		\end{pmatrix}\\
		& =\frac{z_+^{j-j^\prime}}{(z_+ -z_1) (z_+ -z_2)} M(z_+) +
		\frac{z_1^{j-j^\prime}}{(z_1 -z_+) (z_1 -z_2)} M(z_1) +
		\frac{z_2^{j-j^\prime}}{(z_2 -z_+) (z_2 -z_1)} M(z_2)
	\end{aligned}.
\end{equation}
Here, we define $M(z)$ as
\begin{equation}
	M(z) \equiv  \frac{1}{2} \frac{1}{\Delta^2-t^2}  \frac{1}{(z -z_3)(z -z_4)(z-z_-)} \begin{pmatrix} -\mu z-t(z^2+1)  &-\Delta (z^2-1) \\  \Delta (z^2-1) & \mu z+ t (z^2+1)  \end{pmatrix} \otimes \begin{pmatrix} (z+1)^2  & -\frac{V}{J}z(1+z) & -(z+1)^2 
		\\-\frac{V}{J}(z+1)  &\frac{V^2}{J^2}  & \frac{V}{J}(z+1) 
		\\ -(z+1)^2 & \frac{V}{J}z(1+z) & (z+1)^2
	\end{pmatrix}.
\end{equation}
Similarly, for $j<j^\prime$, we substitute $z=e^{-ika}$ find:
\begin{equation}\label{Gpleq}
	\begin{aligned}
		G^p(x_j\leq x_{j^\prime};E=0)
		& = -\ointctrclockwise_{|z|=1} \frac{dz}{2\pi i } \frac{z^{j^\prime-j}}{(z -z_1) (z -z_2) (z-z_+)} M^T(z)\\
		& =-\frac{z_+^{j^\prime-j}}{(z_+ -z_1) (z_+ -z_2)} M^T(z_+) -
		\frac{z_1^{j^\prime-j}}{(z_1 -z_+) (z_1 -z_2)} M^T(z_1) -
		\frac{z_2^{j^\prime-j}}{(z_2 -z_+) (z_2 -z_1)} M^T(z_2)
	\end{aligned}.
\end{equation}
Here, four poles originate from the dispersion of middle two quasiparticle bands:
\begin{equation}
	(z_1,z_2,z_3,z_4) = \left( 
	\frac{-\mu+\sqrt{4(\Delta^2-t^2)+\mu^2}}{2(\Delta+t)},
	\frac{-\mu-\sqrt{4(\Delta^2-t^2)+\mu^2}}{2(\Delta+t)},
	\frac{2(\Delta+t)}{-\mu+\sqrt{4(\Delta^2-t^2)+\mu^2}}, \frac{-2(\Delta+t)}{\mu+\sqrt{4(\Delta^2-t^2)+\mu^2}},
	\right).
\end{equation}
Another two poles result from Bloch eigenvectors $u_{0}(k) u^\dagger_{0}(k)$:
\begin{equation}
	z_\pm = -\left(1+\frac{V^2}{4J^2}\right) \pm \frac{V}{2J} \sqrt{\frac{V^2}{4J^2} +2}
\end{equation}
Among them, only poles $z_1$, $z_2$ and $z_+$ within the integration contour contributes (for positive and real $t$ and $\Delta$, and $|\mu|<2t$). 
By substitute Eq.~(\ref{Gpgeq}) and Eq.~(\ref{Gpleq}) to Eq.~(\ref{eq_MZM_wavefunc}), one can get the MZMs' wavefunctions. Here we take the right Majorana wavefunction as an example.
\begin{equation}\label{psiRx}
	\begin{aligned}
		\psi_{\text{R}}(x) 
		= \left[ \frac{M^T(z_+)}{(z_+ -z_1) (z_+ -z_2)} z_+^{N-j+1} +
		\frac{M^T(z_1)}{(z_1 -z_+) (z_1 -z_2)} z_1^{N-j+1} +
		\frac{M^T(z_2)}{(z_2 -z_+) (z_2 -z_1)} z_2^{N-j+1} \right]  U_{01}^\dagger \psi(Na)
	\end{aligned}
\end{equation}
The localization length of edge states is related to the poles $|z_m|<1$ by the relation $\xi_m = -a/\ln|z_m|$. Therefore, from Eq.~(\ref{psiRx}), there are three decaying modes shows up.  

Specifically, when $\mu=0$, $z_1=-z_2 = i\sqrt{\frac{t-\Delta}{t+\Delta}} >0$ for $t>\Delta$. Wavefunctions in Eq.~(\ref{psiRx}) can be simplified as:
\begin{equation}
	\psi_{\text{R}}(x_j) = A_{\text{R}}^{\text{QM}} \exp\left[-\frac{(N-j)a}{\xi_+}\right] +
	A_{\text{R}}^{\text{BCS}} \exp\left[-\frac{(N-j)a}{\xi_{\text{BCS}}}\right],
\end{equation}
by assuming 
\begin{equation}
	\begin{aligned}
		&A_{\text{R}}^{\text{QM}} = (-1)^{N-j} \frac{z_+ }{z_+^2 -z_1^2}M^T(z_+)  
		U_{01}^\dagger \psi(Na)\\
		&A_{\text{R}}^{\text{BCS}} = \frac{1}{2z_1}\left[\frac{z_1}{z_1 -z_+} M^T(z_1)
		+(-1)^{N-j} \frac{-z_1}{z_1 +z_+}  M^T(-z_1) \right]U_{01}^\dagger \psi(Na)
	\end{aligned}
\end{equation}
Here, $\xi_{\text{BCS}} = -2a/\ln{(\frac{t-\Delta}{t+\Delta})}$ ($t>\Delta>0$)and $\xi_+  =-a/\ln|z_+|$. $A_{\text{R}}$ is the amplitude, which is different for the wavefunction away from the left boundary for the absence of inversion symmetry of the underlying lattice. Additionally, $|A_{\text{R}}^{\text{QM/BCS}} |^2$ give the weights of two modes. It is worth noting that $|A_{\text{R}}^{\text{BCS}}|^2$ are the different for even or odd sites away from two boundaries, which explains the oscillation of MZMs' wavefunctions for the Kitaev mode.  $\xi_+$ goes to the limit $\sqrt{2}a J/V $ to the order of $\mathcal{O}(V/J)$, which is proportional to QML as defined in Eq.~(\ref{eq_qmlength}) to the order of $ \mathcal{O}(V/J)$: 
\begin{equation}
	\xi_+ = 4\xi_{\text{QM}}.
\end{equation}

In particular, when $t=\Delta$, only this decay mode survives. In other words, although the traditional localization length disappears for the MZMs, there still exists the QML as a scale to govern this multi-band system. 

%%%%%%%%%%%%%%%%%%%%%%%%%%%%%%%%%%%%%%%%%%%%%%%%%%%%%%%%%%%%%%%%%%%%%%%%%%%%%%%%%%%%%%%%%%%%%%%%%%%%%%%%%%%%%%%%%%%%%%%%%%%%%%%%%%%%%%%%%%%%%%%%%%%%%%%%%%%%%%%%%%%%%%%%%%%%%%%%%%%%%%%%%%%%%%%%%%%%%%%%%%%%%%%%%%%%%%%%%%%%%%%%%%%%%%%%%%%%%%%%%%%%%%%%%%%%%%%%%%%%%%%%%%%%%%%%%%%%%%%%%%%%%%%%%%%%%%%%%%%%%%%%%%%%%%%%%%%%%%%%%%%%%%%%%%%%%%%%%%%%%%%%%%%%%%%%%%%%%%%%%%%%%%%%%%%%%%%%%%%%%%%%%%%%%%%%%%%%%%%%%%%%%%
\section*{\bf{\uppercase\expandafter{III. Length Dependence of Crossed Andreev Reflection Probability}}}
This section aims to provide additional data regarding the long range crossed Andreev reflections (CARs).
%================================================================================================================================================
\subsection*{I. Method}
In this subsection, we show the recursive Green function method used in the main text in detail. Two semi-infinite normal metal leads are attached to two sides of the topological superconductor. We start with the formulation of the real space tight-binding Hamiltonian associated with  $H_{\text{BdG}}$ in Eq.~(\ref{Eq_BdG_majorana}), and then calculate the scattering matrix of the junction using recursive Green function method. The scattering matrix element at zero temperature is \cite{s_PhysRevLett.47.882, s_PhysRevB.23.6851,s_datta1997electronic}
\begin{equation}
	r_{ij}^{\alpha\beta} = -\delta_{ij}\delta_{\alpha\beta} +i \left[\Gamma_i^\alpha\right]^{1/2} \left[G^R\right]_{\alpha\beta}^{ij} \left[\Gamma_j^\beta\right]^{1/2} .
\end{equation}
Here, $r_{ij}^{\alpha\beta}$ is the scattering matrix element from $\beta$ channel in lead $j$ to $\alpha$ channel in lead $i$, with $i,j = 1$ or $2$ representing left or right lead respectively and $\alpha,\beta \in(e,h)$ denoting two channels. $\left[G^R\right]_{\alpha\beta}^{ij}$ is the matrix element of retarded Green function $G^R$. $\Gamma_i^\alpha = i\left[(\Sigma_i^\alpha)^R -(\Sigma_i^\alpha)^A\right]$, where $(\Sigma_i^\alpha)^{R(A)}$ is the retarded (advanced) self-energy of $\alpha$ particle in lead $i$. 

In the following texts, we use $R_{\alpha\beta} = \Tr \left(r_{11}^{\alpha\beta} (r_{11}^{\alpha\beta} )^\dagger\right)$ to denote reflection probability and $T_{\alpha\beta} = \Tr \left(r_{21}^{\alpha\beta} (r_{21}^{\alpha\beta})^\dagger\right)$ to represent transmission probability. Specifically, $R_{\text{AR}}=R_{he}$ should be the local Andreev reflection (AR) and $T_{\text{CAR}} =T_{he}$ is the crossed Andreev reflection probability.

%================================================================================================================================================
\subsection*{J. Comparison of CAR amplitude verses device length at $E=0$}
\begin{figure}[ht]
	\centering
	\includegraphics[width=1\linewidth]{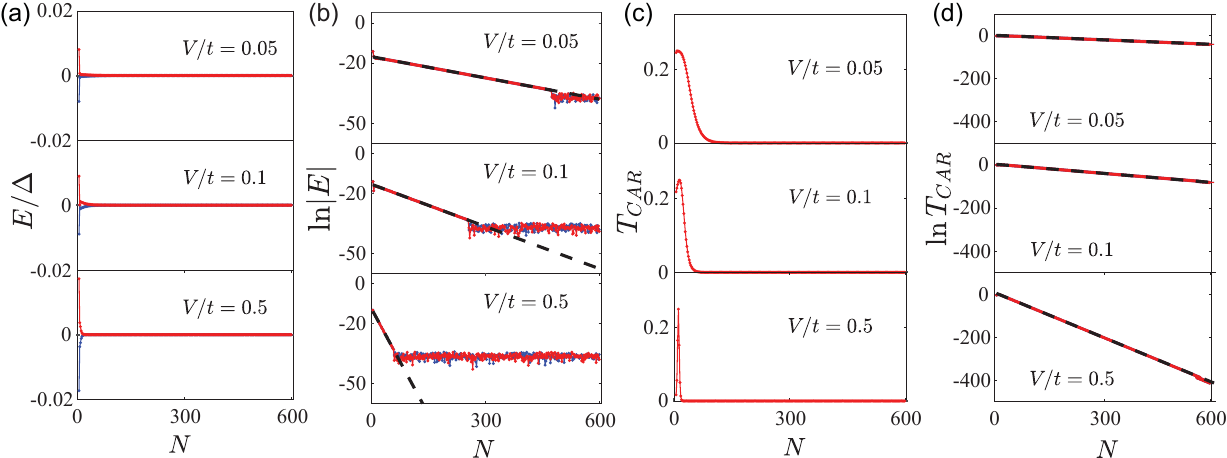}
	\caption{(a) The energy of two Majorana modes verses device length $L=Na$ for different $V/J$. (b) $\ln{|E|}$ verses device length $N$ for different $V/J$. (c) The CAR probability $T_{\text{CAR}}$ as a function of device length $N$ for different $V/J$.  (d)  $\ln{T_{\text{CAR}}}$ verses device length $L=Na$ for different $V/J$. Parameters are $J=1$, $t = 1\times 10^{-4}J$, $\Delta =0.8t$ and $\mu=0.2t$ for all figures. In (a) and (b), blue and red line denote two fermionic modes, the dotted black line represents linear fit. }\label{fig_Sup_expTCAR_ETR}
\end{figure}

In this subsection, we show the detailed extraction of the exponential relationship  $E_\text{M}\propto e^{-Na/(4\xi_{\text{QM}})}$ and $T_{\text{CAR}}\propto e^{-Na/(2\xi_{\text{QM}})}$. In Fig.~\ref{fig_Sup_expTCAR_ETR} (a), we conduct an analysis of the energy of Majorana modes as a function of the device length $N$. The corresponding  $\ln|E|$ in Fig.~\ref{fig_Sup_expTCAR_ETR} (b) reveals an exponential decay of the Majorana mode energies over $N$, with a slower decay observed for smaller values of $V/J$. By performing linear fitting on $\ln(|E|)$ as a function of $N$, we determine the decay length $\xi_{\text{E}}$ to be equal to $4\xi_{\text{QM}}$, confirming the results mentioned in the letter.  Additionally, in Fig.~\ref{fig_Sup_expTCAR_ETR} (c)-(d), we present the CAR probability $T_{\text{CAR}}$ and its logarithm $\ln(T_{\text{CAR}})$ as functions of the device length $N$. These figures clearly indicates an exponentially decaying feature, we determine that the decay length $\xi_T$ is precisely equal to $2\xi_{\text{QM}}$.

%================================================================================================================================================
\subsection*{K. Disorder and finite temperature results}
\begin{figure}[ht]
	\centering
	\includegraphics[width=0.9\linewidth]{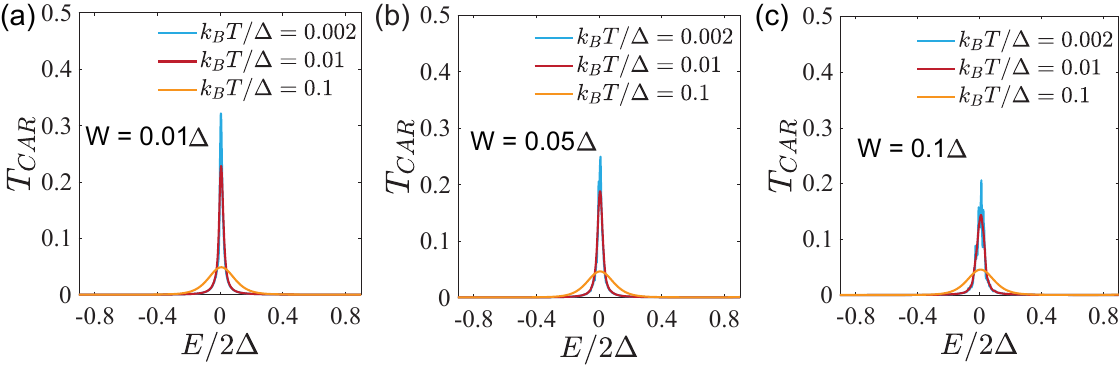}
	\caption{CAR for the Lieb-Kitaev model at different finite temperatures (represented by different colors) with a random disorder strength in $[-W,W]$. The disorder strengths are $W=0.01\Delta$ in (a), $W=0.05\Delta$ in (b) and $W=0.1\Delta$ in (c), respectively.  Parameters are $J=1$, $t = 1\times 10^{-4}J$, $\Delta =0.8t$, $\mu=0.2t$, $V/J=0.01$ and $L = 100a$ for all figures. }\label{fig_Sup_TCAR_disorder}
\end{figure}

In this subsection, we show the robust survival of CAR in the ultra-long sample over random disorder and finite temperature. The CAR probability at different temperature and different disorder strength are plotted in Fig.~\ref{fig_Sup_TCAR_disorder}. The random disorder is simulated by setting and random onsite potential within the energy window $[-W,W]$. Surprisingly, the peak of CAR survives even with the disorder strength comparable to $0.05\Delta$ at the temperature $T= 0.01\Delta/k_B$ (red line in Fig.~\ref{fig_Sup_TCAR_disorder}(b)). 

%%%%%%%%%%%%%%%%%%%%%%%%%%%%%%%%%%%%%%%%%%%%%%%%%%%%%%%%%%%%%%%%%%%%%%%%%%%%%%%%%%%%%%%%%%%%%%%%%%%%%%%%%%%%%%%%%%%%%%%%%%%%%%%%%%%%%%%%%%%%%%%%%%%%%%%%%%%%%%%%%%%%%%%%%%%%%%%%%%%%%%%%%%%%%%%%%%%%%%%%%%%%%%%%%%%%%%%%%%%%%%%%%%%%%%%%%%%%%%%%%%%%%%%%%%%%%%%%%%%%%%%%%%%%%%%%%%%%%%%%%%%%%%%%%%%%%%%%%%%%%%%%%%%%%%%%%%%%%%%%%%%%%%%%%%%%%%%%%%%%%%%%%%%%%%%%%%%%%%%%%%%%%%%%%%%%%%%%%%%%%%%%%%%%%%%%%%%%%%%%
\bibliographystyle{apsrev4-1}

\end{document}